\documentclass{aa}  

\usepackage{graphicx}
\usepackage{txfonts}
\usepackage{multirow}
\usepackage{array}
\newcommand{\PreserveBackslash}[1]{\let\temp=\\#1\let\\=\temp}
\newcolumntype{C}[1]{>{\PreserveBackslash\centering}p{#1}}
\usepackage{subfigure}
\usepackage{hyperref}
\usepackage{xcolor}

\begin{document}

   \title{Binary microlensing by high eccentric stellar-mass black hole binaries}

   \subtitle{}

   \author{Kyungmin Kim\inst{1} 
          \and 
          Yeong-Bok Bae\inst{2}\fnmsep\thanks{Corresponding author} 
          \and 
          Yoon-Hyun Ryu\inst{1}
          }

   \institute{Korea Astronomy and Space Science Institute, 776 
              Daedeokdae-ro, Yuseong-gu, Daejeon 34055, Republic of Korea\\
              \email{kkim@kasi.re.kr}
        \and 
              Department of Physics, Chung-Ang University, 84 Heukseok-ro, Dongjak-gu, Seoul 06974, Republic of Korea\\
              \email{astrobyb@gmail.com}
             }
          
   \date{Received ; accepted }

 
  \abstract
   {Microlensing is one of the most promising tools for discovering stellar-mass black holes (BHs) in the Milky Way because it allows us to probe dark or faint celestial compact objects.}
   {While the existence of stellar-mass BHs has been confirmed through observation of X-ray binaries within our galaxy and gravitational waves from extragalactic BH binaries, a conclusive observation of microlensing events caused by Galactic BH binaries has yet to be achieved. In this study, we focus on those with high eccentricity, including unbound orbits, which can dynamically form in star clusters and could potentially increase the observation rate.}
   {We demonstrate parameter estimation for simulated light curves supposing various orbital configurations of BH binary lenses. We employ a model-based fitting using the Nelder-Mead method and Bayesian inference based on the Markov chain Monte Carlo method for the demonstration.}
   {The results show that we can retrieve true values of the parameters of high eccentric BH binary lenses within the $1\sigma$ uncertainty of inferred values.}
   {We conclude it is feasible to find high eccentric Galactic BH binaries from the observation of binary microlensing events.}


   \maketitle
%

\section{Introduction}

Microlensing is a powerful tool that allows us to probe dark or faint celestial objects, such as planets, dwarfs, neutron stars, and black holes (BHs), as well as massive compact halo objects by magnifying the luminosity of background sources~\citep{Paczynski:1986, Mao:1991, Gould:1992}. Thanks to several photometric surveys conducted hitherto—such as Exp\'{e}rience pour la Recherche d'Objets Sombres (EROS), MAssive Compact Halo Objects (MACHO) project, Microlensing Observations in Astrophysics (MOA), Optical Gravitational Lensing Experiment (OGLE), and Korea Microlensing Telescope Network (KMTNet)---tens of thousands of microlensing events within our own and nearby galaxies have been observed~\citep{Aubourg:1993, Alcock:2000, Sumi:2011, Udalski:2015, Yock:2018, Byun:2022}.

Stellar-mass BHs are one of the possible lens objects of microlensing events. They have been identified through X-ray observations ~\citep{Liu:2007, Tetarenko:2016}, but microlensing provides a unique opportunity to detect isolated stellar-mass BHs as well. Based on the stellar mass function, stellar-mass BHs are expected to account for approximately 1\% of microlensing events ~\citep{Gould:1999xx}. However, only about a dozen stellar-mass BH candidates have been found from microlensing events to date, even though they are anticipated to be overrepresented in the observation due to the longer event duration resulting from their substantial masses ~\citep{Agol:2002, Bennett:2002, Mao:2002, Poindexter:2005, Dong:2007, Shvartzvald:2015, Wyrzykowski:2016, Mroz:2021, Sahu:2022}.

As for stellar-mass BH binaries, they have not been conclusively discovered through microlensing events, which contrasts sharply with the fact that nearly 100 BH binaries have been identified through gravitational-wave (GW) observations ~\citep{LIGOScientific:2018mvr, LIGOScientific:2020ibl, LIGOScientific:2021usb, LIGOScientific:2021djp}. However, we should note that this situation arises from differences in the observable spatial volume and time, rather than suggesting a scarcity of wide-orbit BH binaries that produce microlensing events. This lack of detection could be attributed to the small microlensing parallax of the BH binary due to their large mass as well as the lens’s orbital motion that complicates the evaluation of the parallax, which makes it difficult to accurately determine the lens mass~\citep{Karolinski:2020, Ma:2021ucq}.

It is known that BH binaries can form through the evolution of massive stellar binaries \citep{Marchant:2023, Langer:2020}. Also, they can form via dynamical interactions at the center of star clusters \citep{OLeary:2009, Banerjee:2010, Bae:2014, Hong:2015, Park:2017}. In particular, in highly dense regions like the center of nuclear star clusters, many BHs can be gathered in a small volume through the dynamical friction and mass segregation \citep{Hailey:2018}, so the number of interactions of BHs increases, which triggers the dynamical formation of BH binaries.

In this study, we consider BH binary lenses with a range of eccentricities, from low to high, including those in unbound orbits with eccentricities equal to or exceeding unity. Generally, the formation of BH binaries from massive stellar binaries tends to favor low eccentricity, while dynamical formation processes markedly favor the production of BH binaries with high eccentricities~\citep{Bae:2014, Hong:2015}. Furthermore, gravitational perturbations from the supermassive BH at the center of the galaxy can amplify the eccentricity of BH binaries, known as the Kozai-Lidov mechanism~\citep{Hoang:2018}. Additionally, in multi-body interactions—such as binary-single or binary-binary encounters—highly eccentric encounters occur repeatedly ~\citep{Samsing:2018, Zevin:2019}, commonly resulting in unbound orbits. Thus, the occurrence rate of high-eccentricity encounters, especially in cluster centers, is significant and cannot be overlooked.

In the determination of the characteristics of binary microlensing from a light curve of a luminous celestial source, well-aligned configurations between the source, binary lens, and observer ought to be satisfied, and several fundamental parameters, e.g., the distances to the source and lens, the mass of the binary lens, pericenter distance or separation between the components of a binary, the trajectory and relative velocity of the source, and so on, should be taken into account. Assuming negligible parallax for BH binary lens due to its small value as previously mentioned, in this study, we focus on the effects of orbital properties of BH binary lenses, specifically including unbound BH trajectories, that can appear on the light curve of microlensing events. We explore the signature of binary lensing depending on the eccentricity of a BH binary and compare the cases of elliptic and hyperbolic orbits with varying other parameters characterising a binary lens system, such as the pericentre distance and mass ratio between two component BHs, and the inclination angle between the orbital axis and the line of sight. Additionally, we test whether we can restore the true value of the parameters of a BH binary lens by using a model-based fitting method, the Nelder-Mead fitting~\citep{Nelder-Mead}, and Bayesian inference using the Markov chain Monte Carlo sampling method~\citep{mcmc}.

We find from this study that the employed parameter estimation methods can retrieve true values of the binary microlensing parameters within the $1\sigma$ uncertainty of inferred values. For some samples of which the Nelder-Mead fitting failed in the uncertainty estimation, Bayesian inference based on the Markov chain Monte Carlo method can succeed in estimating the uncertainty. On the other hand, if the median value of the posterior distribution obtained by the Bayesian inference deviates from the true value more than that from the initial Nelder-Mead fitting, it turns out that using the uncertainty estimated from the Bayesian inference helps reduce overall deviations of all parameters.

This paper is organized as follows: In Sec.~\ref{sec:methods}, we summarize how to compute the orbits of two black holes and how to simulate binary microlensing phenomena that can occur by the two black holes acting like a binary lens. We investigate the parameter dependency in the binary microlensing signature and show the resulting caustics and light curve in Sec.~\ref{sec:cri_cau_lc}. We provide the result of parameter estimation for the light curve of a simulated binary microlensing event in Sec.~\ref{sec:pe}. In Sec.~\ref{sec:discussions}, we discuss our result and the future prospect of observing binary microlensing events by stellar-mass binary black holes within our and/or nearby galaxies.

\begin{figure}
    \includegraphics[width=1.\linewidth]{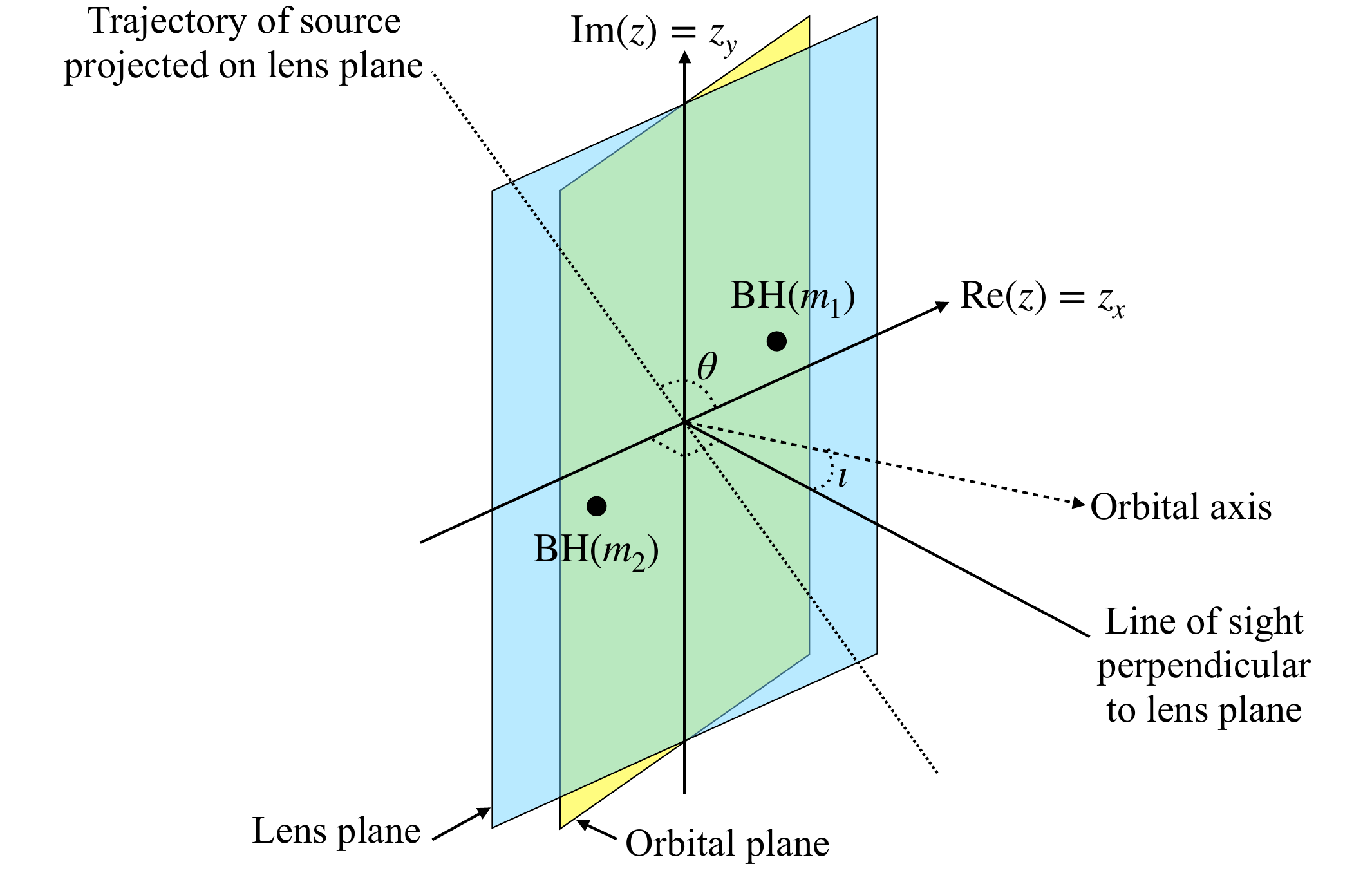}
    \caption{Schematic of BH binary lens system. We suppose the line of sight towards the lens system is perpendicular to the lens plane which may not be identical to the orbital plane of the orbiting two BHs. We define the angle between the orbital axis and the line of sight as the inclination angle $\iota$. For a source constantly moving along the grey dotted line, the angle between the real axis $z_x$ on the lens plane and the source trajectory projected onto the lens plane is defined as $\theta$. \label{fig:schematic}}
\end{figure}

\section{Methods}
\label{sec:methods}

\subsection{Lens equation and magnification for binary lens system}

We consider the light of a luminous source being lensed by two point-like stellar-mass black holes (BHs) acting like a binary lens. As depicted in Fig.~\ref{fig:schematic}, we also take into account the lens plane may not be identical to the orbital plane of the BH binary.

To simulate the binary microlensing by two BHs projected onto the lens plane, we define the complex coordinate axes $z_x$ and $z_y$ on the lens plane corresponding to the real and imaginary parts of a complex coordinate $z$, respectively. Based on the complex coordinate, we then adopt the complex notation of the lens equation~\citep{Witt:1990,Witt:1995,Meneghetti:2021}:
\begin{equation}
z_s = z - \frac{m_1}{z^*-z_1^*} - \frac{m_2}{z^*-z_2^*}~, \label{eq:lens_eq}
\end{equation}
where $z$ is a position on the lens plane, $z_s$ is the position of the source on the source plane, and $z_{1,2}$ are the positions of two lenses on the lens plane. The asterisk ($^*$) denotes the complex conjugate. With the complex conjugate form of Eq.~\eqref{eq:lens_eq}, the lens equation then can be rewritten as a complex polynomial equation of degree 5:
\begin{equation}
c_5 z^5 + c_4 z^4 + c_3 z^3 + c_2 z^2 + c_1 z + c_0= 0~, \label{eq:lens_eq_poly}
\end{equation}
where $c_i(i\!=\!0,1,\ldots,5)$ are the polynomial coefficients and their exact forms are given in Eq.~\eqref{eq:le_poly_coeffs}.

Similar to the lens equation, for a given phase angle $\phi \in [0,2\pi)$ with respect to the real axis of the lens plane, we can obtain the critical curve of the binary lens via
\begin{equation}
\frac{m_1}{(z^*-z_1^*)^2} + \frac{m_2}{(z^*-z_2^*)^2} = e^{i\phi}~, \label{eq:cri_curve}
\end{equation}
and it can be turned into another complex polynomial equation of degree 4:
\begin{equation}
d_4 z^4 + d_3 z^3 + d_2 z^2 + d_1 z + d_0 = 0~, \label{eq:cri_curve_poly}
\end{equation}
with $d_j(j\!=\!0,1,\ldots,4)$ given in Eq.~\eqref{eq:ce_poly_coeffs}. Also, we can compute the caustics on the source plane by inserting the solution of Eq.~\eqref{eq:cri_curve_poly} into Eq.~\eqref{eq:lens_eq}.

The total magnification $\mu$ occurring by the binary lens at a certain observation time $t$ can be obtained as follows
\begin{equation}
\mu(t) = \sum^{N_\textrm{images}}_j |\mu_j(t)|~,
\end{equation}
where $\mu_j$ is the individual magnification of each image $j$ of a source at the given $t$: For a given binary lens system, we can calculate $\mu_j(t)$ by
\begin{equation}
\mu_j(t) = \left(1 - \left| \frac{m_1}{z^*_j(t) - z^*_1} + \frac{m_2}{z^*_j(t) - z^*_2} \right|^2 \right)^{-1}~.
\end{equation}

\subsection{Orbital motion of black hole binary}

As we suppose two BHs causing binary lensing orbit each other with a non-zero eccentricity ($e$), we have to calculate the time-dependent trajectory of two BHs, i.e., $z_1(t)$ and $z_2(t)$, in order to solve Eqs.~\eqref{eq:lens_eq_poly} and \eqref{eq:cri_curve_poly}. For the orbital motion of the two BHs in the orbital plane, we solve the Newtonian equation of motion using the Verlet integration method~\citep{Verlet:1967}, which is a symplectic method that preserves the energy of the system. The time step is fixed at 0.1 day. The resulting solution is a Kepler orbit as is well known. Defining the angle between the orbital axis and the line of sight perpendicular to the lens plane as the inclination angle ($\iota$), the initial condition is adjusted with the given eccentricity and the pericentre distance projected onto the lens plane.


\section{Caustics and light curves}
\label{sec:cri_cau_lc}

In this work, we assume observing a binary lensing event is made with a constant time interval for a total observation time. Following the assumption, we interpolate the trajectory of orbit obtained by solving the equation of motion for the given observation time so that the time steps of the trajectory are equally spaced. Consequently, we can simulate a binary lensing light curve with $\sim 1~\textrm{day}$ cadence. In addition, in order to ensure a sufficient observation time for identifying the signature of binary lensing, i.e., two sharp spikes connected with magnified `U'-shape light curve, we take twice of the Einstein cross time, calculable for the Einstein radius of the total mass of a BH binary, as the total observation time. 

Regarding the consideration discussed thus far and denoting a model parameter of this work as $\lambda$, we configure a set ${\Lambda}$ as follows
\begin{equation}
{\Lambda} = \{\lambda\} = \{D_S, D_L, v_r, \theta, m_1, m_2, s, e, \iota\}~,
\end{equation}
where $D_S$ and $D_L$ are the distance to source and BH binary lens, $v_r$ is the relative velocity of source against to the lens, $\theta$ is the angle between the real axis $z_x$ on the lens plane and the source trajectory projected onto the lens plane, $m_1$ and $m_2$ are the mass of component BHs of a binary lens, $s$ is the pericenter distance between two BHs, $e$ is the eccentricity of the orbital motion of BH binary, and $\iota$ is the inclination angle defined by the angle between the orbital axis and the line of sight. In order to focus on the original and projected orbital motion of binary BHs, we fix $D_S=8~\textrm{kpc}$, $D_L=5~\textrm{kpc}$, $v_r=200~\textrm{km/s}$, $\theta=120^\circ$, and alter $m_1$, $m_2$, $s$, $e$, and $\iota$.

\begin{figure*}
    \centering
    \includegraphics[width=.7\linewidth]{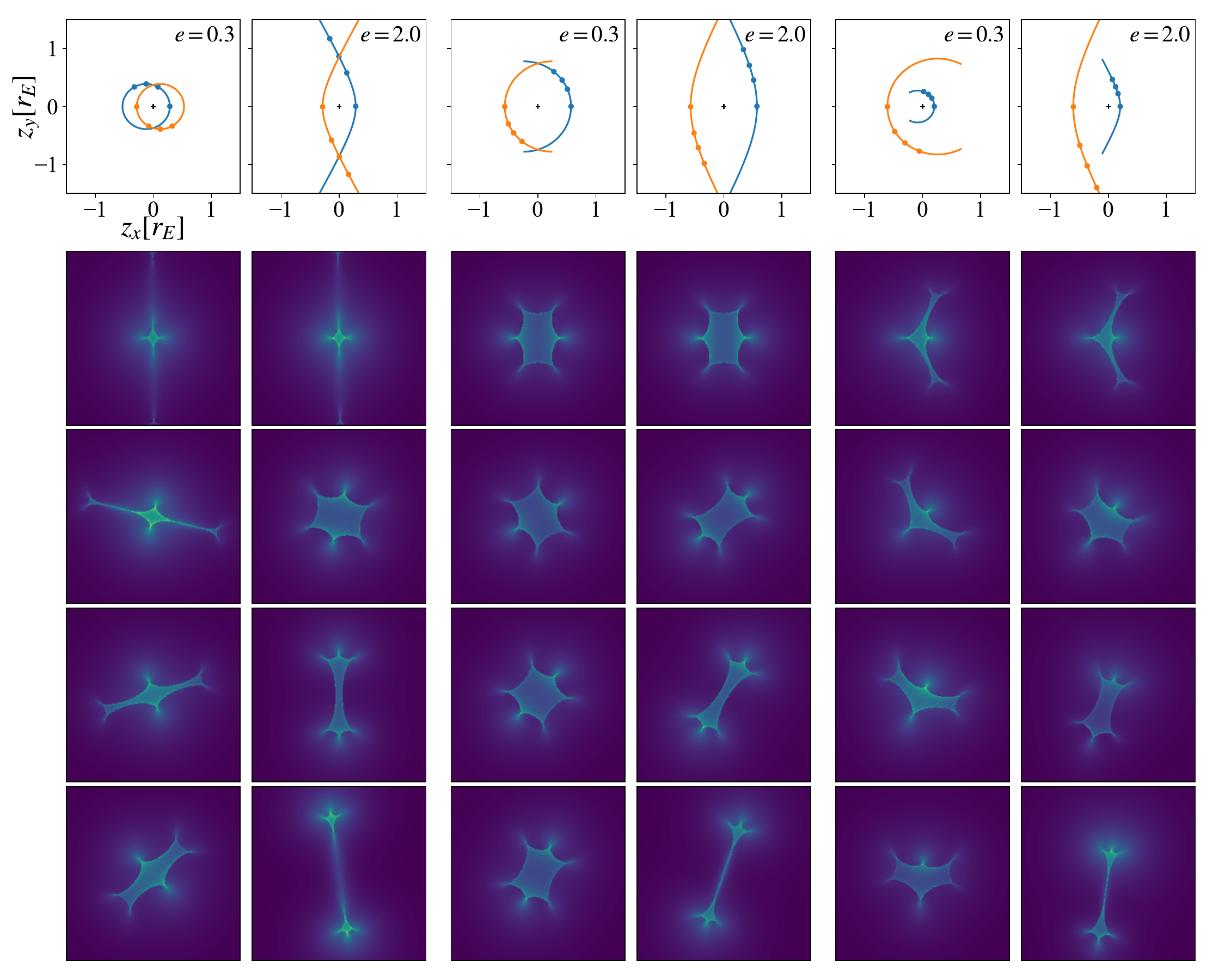}
    \caption{The lens orbits and the magnitude color maps for the selected points on the orbit. The left two columns are for the mass ratio $q=1$ and the pericenter distance $s=10$ AU in the case of elliptic ($e=0.3$) and hyperbolic ($e=2.0$) lens orbits. The middle and right two columns show the cases of $q=1$, $s=20$ AU,  and $q=0.3$ ($m_{1}=30 M_{\odot}$, $m_{2}=10 M_{\odot}$), $s=20$ AU, respectively. The points on the orbits in the left, middle, and right two columns are all at the same time, and the color maps are arranged in the order in which the lens moves away from the pericenter. The positions are normalized to the Einstein radius.}
    \label{fig:colormap}
\end{figure*}

\begin{figure*}
    \centering
    \includegraphics[width=.9\linewidth]{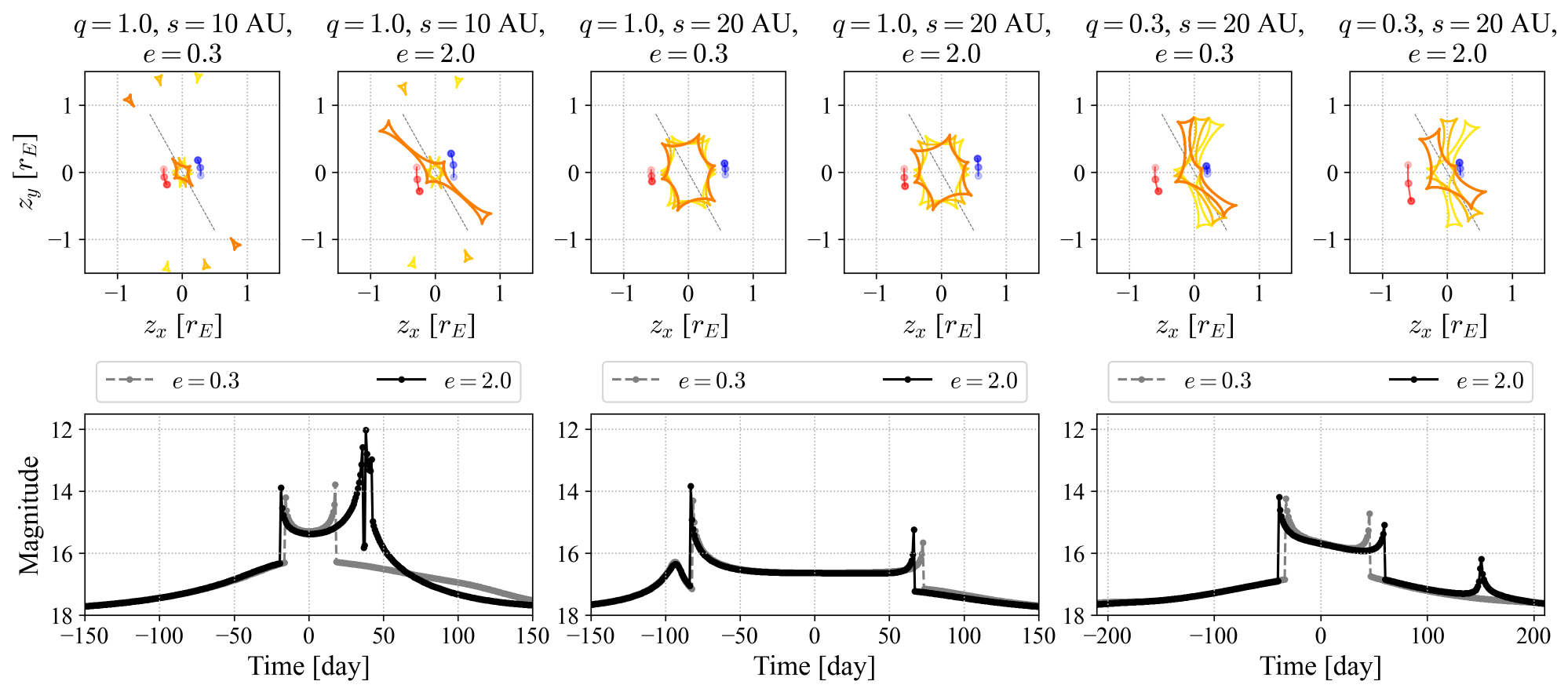}
    \caption{Caustics and corresponding light curve for each case of Fig.~\ref{fig:colormap}. We suppose observing  microlensed light of an arbitrary moving luminous source is made for the twice of the Einstein cross time. For the same given source, we can see that the signature of binary microlensing appearing in the light curve changes by different values of $q$, $s$, and $e$ as expected. \label{fig:caustics_lcs}}
\end{figure*}

\subsection{Zero inclination angle \label{sec:zero_iota}}

To examine the $e$-dependent variations in microlensing signature, e.g., caustics and magnification, we draw color maps depicting the magnification of the source's brightness for five specific lens positions along the lenses' orbits (Fig.~\ref{fig:colormap}). Here, we choose two $e$ values, $e\!=\!0.3$ and $2.0$, as representing values of elliptic and hyperbolic orbits, respectively. Despite of many other parameters contribute to determine the binary microlensing signature as discussed in the previous section, for simplicity, we first consider $\iota=0^\circ$ and assume a source of an arbitrary magnitude in the direction of the Galactic center, with a distance of 8 kpc to the source ($D_{S}=8$ kpc) and 5 kpc to the lens ($D_{L}=5$ kpc). Then, we investigate three example cases of \{$q,~s$\}, i.e., \{$q\!=\!1.0$, $s\!=\!10~\textrm{AU}$\}, \{$q\!=\!1.0$, $s\!=\!20~\textrm{AU}$\}, and \{$q\!=\!0.3$, $s\!=\!20~\textrm{AU}$\}. Hereafter, we label each pair of \{$q,~s$\} as Case I, II, and III, respectively, for convenience.

As is well known, caustics are formed in the direction perpendicular to the connecting direction of the two lenses when the distance between two lenses is close, but as the two lenses move away from each other, the shape of the caustics is formed in the direction connecting the two lenses. As shown in Fig.~\ref{fig:colormap}, in the case of a binary lens with an elliptical orbit, the distance between the two lenses is kept within a certain range, so a caustic of a certain size is also maintained. On the other hand, in the case of a binary lens with a hyperbolic orbit, the caustic becomes elongated as it moves away from the pericenter, and in this case, the magnification of the source brightness is likely to occur for a very short period as the source traverses the caustic. After that, the effect of each point mass lens will occur as the distance becomes larger. 

If the position of lens does not change significantly during the observation period, the brightness change of the source due to its movement can be inferred from this color map. However, in reality, both the lens and the source are moving, so it is difficult to grasp the actual light curve with a magnitude color map at a specific time such as Fig.~\ref{fig:colormap}. In addition, the parameter space that determines the lens effect is very wide, so it is not possible to examine all situations. Thus, we have examined a few examples focused on the parameters that determine the lens dynamics. 

\begin{figure*}
\centering
    \includegraphics[width=.9\linewidth]{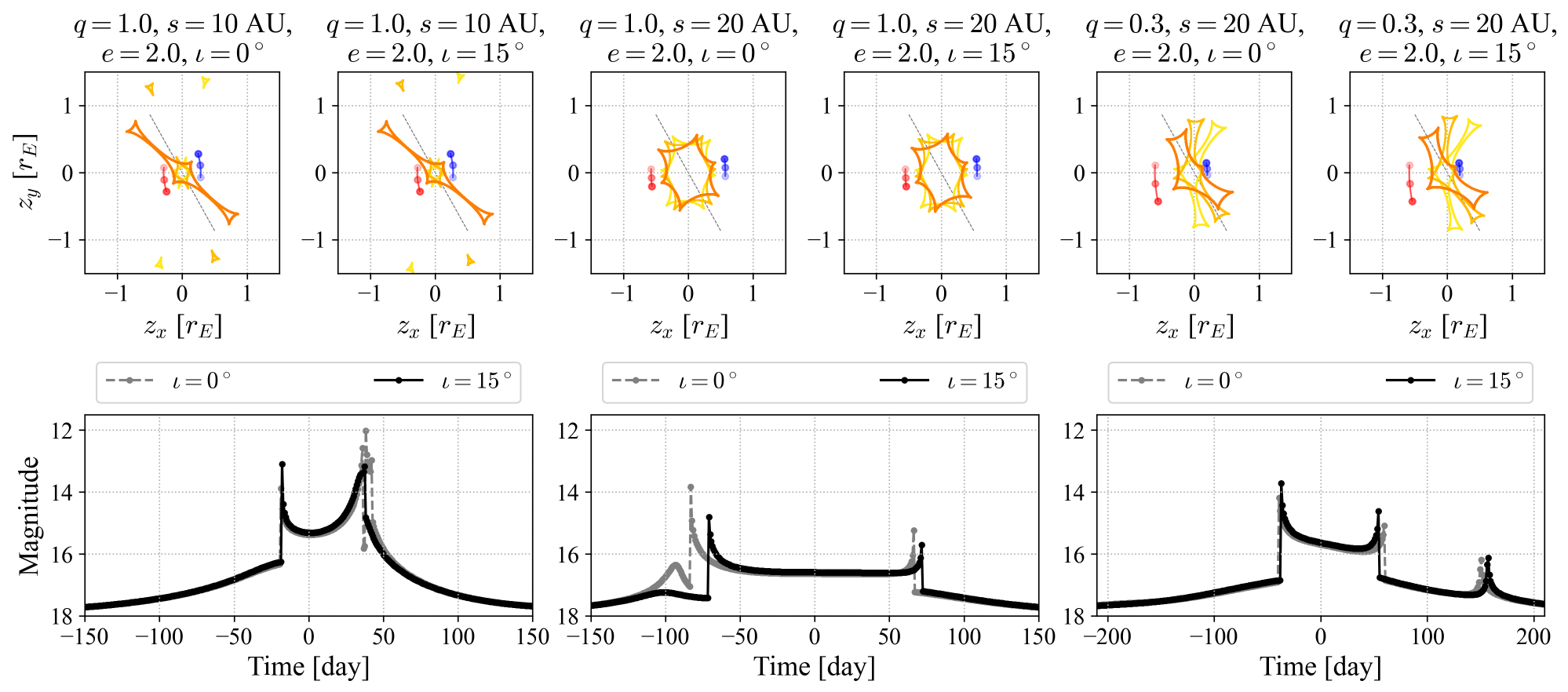}
    \caption{Comparing caustics and corresponding light curves between two $\iota$ values, $\iota=0^\circ$ and $\iota=15^\circ$ for each case in Fig.~\ref{fig:colormap}. One can see that non-zero $\iota$ results in slightly different light curves from those of $\iota=0^\circ$ for the given $q$, $s$, and $e$. \label{fig:caustics_lcs_nonzero_iota}}
\end{figure*}

We explore how the $e$-dependant shapes, sizes, and orientations of caustics shown in Fig.~\ref{fig:colormap} affects the profile of light curves. To this end, we simulate light curves of the same source and BH binary lenses considered in Fig.~\ref{fig:colormap}. We further suppose the source moves along a straight path corresponding to $\theta=120^\circ$ with a constant relative velocity $v_r=200~\textrm{km/s}$ to the BH binary. We set the observation period as the twice of the Einstein cross time calculated by the given binary microlensing configuration. For simplicity, we assume the background source passes the barycentre of the BH binary, i.e., zero impact parameter, at the middle of the observation period.

Fig.~\ref{fig:caustics_lcs} shows the locations and orientations of caustics for given $q$, $s$, and $e$ (top panels) and corresponding light curves (bottom panels). We can see that the $e$-dependent differences in caustics determine not only the width and height of `U'-shape double peaks and the presence of another double peak or single peak on the light curve. It is easy to understand that such differences originate from the changes in the shape, size, and orientation of caustics for the given observation period and the location of the background source at each moment of observation.

We have compared the $e$-dependent characteristics of caustics between $e=0.3$ and $e=2.0$ corresponding to an elliptic and hyperbolic orbits, respectively, and resultant profiles of binary microlensing light curves thus far. We have indeed seen that a BH binary of a hyperbolic orbit results in different light curves compared to that of an elliptic orbit: Amongst tested cases, light curves of Case I or Case III of $e=2.0$ show wider `U'-shape double peaks and another peak(s) after the first appearance of `U'-shape peaks. However, Case II of $e=2.0$ only shows small difference in the width and height of the `U'-shape peaks compared to that of $e=0.3$ because of relatively smaller change in the size and orientation of the caustics.

\subsection{Non-zero inclincation angle}

Next, we consider $\iota\neq0^\circ$ because it is more realistic than $\iota=0^\circ$. We test how much non-zero $\iota$ changes the caustics and light curve because, if $\iota \neq 0^\circ$, $s$ and $e$ of two BHs projected on the source plane will be obviously adjusted by $\cos \iota$ from their actual values. In Fig.~\ref{fig:caustics_lcs_nonzero_iota}, we present two cases of $\iota=0^\circ$ and $15^\circ$ for the configurations studied in Figs.~\ref{fig:colormap} and \ref{fig:caustics_lcs}, but focusing on $e=2.0$. 

One can see that, if $\iota=15^\circ$, not only the projected $s$ of two BHs becomes shorter than that of $\iota=0^\circ$ but also the shape and orientation of the caustics varies for both $e=0.3$ and $e=2.0$ as expected. For instance, the caustics of $\iota=15^\circ$ is slightly narrower than $\iota=0^\circ$ for all cases because of the shorten $s$ and increased $e$ on the source plane. Also, we can observe from the light curve for each $\iota$ that the position and number of spikes shown when $e=2.0$ are different from the spikes of $e=0.3$ because of the $e$-dependent variations in the shape, size, and orientation of corresponding caustics.

Thus far, we have investigated the difference between $e=0.3$ and $e=2.0$ via example caustics and light curves. We have seen from the examples presented in the previous section that the shape and orientation of the caustic of two stellar-mass BHs indeed depends on the tested parameters and they consequently affect the profile of the light curve. In specific, we could see that $e=2.0$ gives different characteristics of a BH binary lensing compared to the case of $e=0.3$. However, we understand from this investigation that it is hard to find a general pattern in considering $e=2.0$ instead of $e=0.3$ because caustics and resultant light curve are highly degenerated. For example, when we compare $e=0.3$ and $2.0$ with $\iota=0^\circ$, we could see additional peak(s) from Case I and III, but, from Case II, we only saw small changes in the width and height of the `U'-shape double peaks. Similarly, if we consider $\iota\neq0^\circ$, it is also nontrivial to find a common behaviour for $e=2.0$ because, for example, the secondary `U'-shape double peaks of Case I and a single bump around $-100$ day of Case II that are shown in $\iota=0^\circ$ disappear when $\iota=15^\circ$. On top of that, there is quite small shift in the location and three peaks of Case III for $\iota=15^\circ$ that makes us be confused in recognising the signature of $e=2.0$ from a simple visual inspection.

This can be a limitation on finding a general feature about the effect of different eccentricities of a BH binary lens system. Moreover, if we consider other degrees of freedom such as the source trajectory, distance to the binary lens, and distance to the source, the degeneracy will be more complicated. At this point, the discussed limitation raises a question whether we can correctly recover those highly degenerated parameters of a BH binary lens system if a light curve represents the signature of binary microlensing. To answer the question, in the following section, we demonstrate parameter estimations in order to investigate whether the unknown characteristic parameters of a BH binary lens system can be correctly recovered from a binary microlensing light curve.

\section{Parameter Estimation}
\label{sec:pe}

We demonstrate a fitting-based parameter estimation of binary microlensing events. In practice, a successful fitting of a given light curve data with a binary microlensing model can be a primary approach for estimating the most promising values of microlensing parameters from the value of model parameters used in the fitting. We adopt the Nelder-Mead (NM) method~\citep{Nelder-Mead} as the fitting method of this work.

\begin{table}
\caption{Prior ranges for the Nelder-Mead fitting. We assume uniform distribution for all priors for simplicity.}
\label{tab:prior}
\centering
\begin{tabular}{l c c}
\hline
{Parameter (unit)} & {Prior range} & Distribution \\
\hline
$D_S$ (kpc) & $[0,30]$ & Uniform \\
$D_L$ (kpc) & $[0,15]$ & Uniform \\
$v_r$ (km/s) & $[150, 250]$ & Uniform \\
$\theta$ ($^\circ$) & $[0,360]$ & Uniform \\
$m_1$ ($M_\odot$) & $[5,100]$ & Uniform \\
$m_2$ ($M_\odot$) & $[5,100]$ & Uniform \\
$s$ (AU) & $[1,50]$ & Uniform \\
$e$ (--) & $[0,10]$ & Uniform \\
$\iota$ ($^\circ$) & $[-90,+90]$ & Uniform \\
\hline
\end{tabular}
\end{table}

\begin{table*}
\addtolength{\tabcolsep}{-0.2em}
\scriptsize
\caption{Estimated parameters from the Nelder-Mead fitting method for samples of test parameters, $m_1$, $m_2$, $s$, and $e$, including parameters Case I, II, and III of Figs.~\ref{fig:caustics_lcs} and \ref{fig:caustics_lcs_nonzero_iota}. Samples 1--15, 16--30, and 31--45 fundamentally correspond to Case I, II, and III, respectively, but more values of $e$ and $\iota$ are tested. The number in parentheses indicates estimated uncertainty $\sigma_\lambda$ of each parameter $\lambda$, if available. We summarize $\chi^2_r$ to see the fitting quality and $\left<\delta_\lambda\right>$ to evaluate the performance of the fitting-based parameter estimation, respectively, in the last two columns. Note that we have discarded samples of $\iota=0^\circ$ because not only it is less realistic but also $\iota_T=0^\circ$ makes $\delta_{\iota}$ diverge. \label{tab:nm_fitting}}
\centering
\begin{tabular}{@{} c c c c c C{0.01cm} c c c c c C{0.01cm} c c c @{}}
\hline
\multirow{3}{*}{\begin{minipage}{0.7cm}
\centering Sample no.\end{minipage}} & \multicolumn{4}{c}{Fixed parameters} & & \multicolumn{5}{c}{Altered parameters} & & \multicolumn{2}{c}{Fitting results} & \multirow{3}{*}{Remarks} \\
\cline{2-5} \cline{7-11} \cline{13-14}
\\[-1em]
& $D_S$ & $D_L$ & $v_r$ & $\theta$ & & $m_1$ & $m_2$ & $s$ & $e$ & $\iota$ & & $\chi^2_r$ & $\left<\delta_\lambda\right>$ & \\
& (kpc) & (kpc) & (km/s) & ($^\circ$) & & ($M_\odot$) & ($M_\odot$) & (AU) & (-) & ($^\circ$) & & (-) & (-) & \\
\\[-1em]
\hline
\\[-1em]
1 & $8.01(\textrm{N/A})$ & $5.00(\textrm{N/A})$ & $200.01(\textrm{N/A})$ & $119.92(\textrm{N/A})$ & & $9.99(\textrm{N/A})$ & $10.00(\textrm{N/A})$ & $10.00(\textrm{N/A})$ & $0.98(\textrm{N/A})$ & $5.03(\textrm{N/A})$ & & $1.14$ & $2.75\!\times\!10^{-3}$ & \\
2 & $7.72(0.26)$ & $5.01(0.49)$ & $199.99(17.68)$ & $119.94(0.08)$ & & $10.59(1.49)$ & $10.58(1.49)$ & $9.94(0.62)$ & $0.90(0.18)$ & $14.58(9.34)$ & & $0.99$ & $3.26\!\times\!10^{-2}$ & \\
3 & $8.22(0.01)$ & $5.01(0.02)$ & $201.08(7.44)$ & $120.03(0.09)$ & & $9.63(0.03)$ & $9.63(0.03)$ & $10.03(0.01)$ & $1.10(0.14)$ & $29.99(0.17)$ & & $1.08$ & $2.24\!\times\!10^{-2}$ & \\
4 & $7.96(0.70)$ & $5.03(0.97)$ & $200.00(6.67)$ & $120.00(0.08)$ & & $10.08(1.48)$ & $10.07(1.48)$ & $9.98(0.82)$ & $1.48(0.08)$ & $4.81(9.54)$ & & $1.05$ & $8.50\!\times\!10^{-3}$ & \\
5 & $8.07(0.54)$ & $4.99(0.80)$ & $200.00(14.15)$ & $120.06(0.08)$ & & $9.91(1.36)$ & $9.91(1.36)$ & $10.03(0.66)$ & $1.55(0.18)$ & $14.99(7.58)$ & & $1.07$ & $7.67\!\times\!10^{-3}$ & \\
6 & $8.03(0.02)$ & $4.95(0.05)$ & $200.00(5.57)$ & $120.07(0.08)$ & & $10.10(0.11)$ & $10.10(0.11)$ & $9.99(0.03)$ & $1.48(0.12)$ & $28.72(0.50)$ & & $1.03$ & $9.85\!\times\!10^{-3}$ & \\
7 & $8.04(0.34)$ & $5.11(0.77)$ & $200.32(25.33)$ & $120.01(0.08)$ & & $9.97(2.00)$ & $9.97(2.00)$ & $9.99(0.88)$ & $2.02(0.37)$ & $6.89(30.88)$ & & $1.05$ & $4.69\!\times\!10^{-2}$ & 2nd largest $\left<\delta_\lambda\right>$ \\
8 & $8.00(0.25)$ & $4.99(0.64)$ & $200.01(25.29)$ & $119.94(0.07)$ & & $10.01(1.78)$ & $9.99(1.78)$ & $10.00(0.70)$ & $1.99(0.37)$ & $15.01(13.90)$ & & $0.90$ & $1.22\!\times\!10^{-3}$ & \\
9 & $8.30(0.04)$ & $4.98(0.10)$ & $200.68(6.77)$ & $119.99(0.07)$ & & $9.91(0.20)$ & $9.94(0.20)$ & $10.34(0.08)$ & $2.16(0.19)$ & $30.68(0.81)$ & & $0.99$ & $2.20\!\times\!10^{-2}$ & \\
10 & $7.95(0.34)$ & $5.09(0.78)$ & $199.93(29.25)$ & $119.89(0.08)$ & & $9.94(2.19)$ & $9.94(2.19)$ & $9.97(0.92)$ & $2.48(0.49)$ & $10.00(24.39)$ & & $1.09$ & $1.17\!\times\!10^{-1}$ & Largest $\left<\delta_\lambda\right>$ \\
11 & $7.77(0.37)$ & $4.96(0.68)$ & $200.04(21.53)$ & $120.07(0.08)$ & & $10.48(1.89)$ & $10.50(1.89)$ & $10.05(0.81)$ & $2.37(0.36)$ & $15.54(11.19)$ & & $1.08$ & $2.55\!\times\!10^{-2}$ & \\
12 & $7.95(0.06)$ & $5.00(0.13)$ & $199.99(8.32)$ & $119.98(0.08)$ & & $10.14(0.36)$ & $10.15(0.36)$ & $9.99(0.14)$ & $2.44(0.23)$ & $29.68(1.35)$ & & $1.16$ & $7.84\!\times\!10^{-3}$ & \\
13 & $8.01(10^{-3})$ & $5.00(0.01)$ & $200.00(4.32)$ & $119.93(0.10)$ & & $9.98(0.02)$ & $9.99(0.02)$ & $10.00(10^{-3})$ & $2.97(0.14)$ & $4.99(0.55)$ & & $1.04$ & $1.74\!\times\!10^{-3}$ & \\
14 & $8.00(0.60)$ & $5.00(0.89)$ & $200.00(16.60)$ & $120.03(0.08)$ & & $9.98(1.60)$ & $10.01(1.60)$ & $10.00(0.78)$ & $3.00(0.34)$ & $15.04(8.79)$ & & $1.06$ & $6.86\!\times\!10^{-4}$ & \\
15 & $7.88(0.07)$ & $5.00(0.15)$ & $199.96(8.52)$ & $120.07(0.07)$ & & $10.28(0.45)$ & $10.27(0.45)$ & $9.95(0.17)$ & $2.89(0.25)$ & $29.40(1.63)$ & & $0.85$ & $1.50\!\times\!10^{-2}$ & \\
\\[-1em]
\hline
\\[-1em]
16 & $8.00(\textrm{N/A})$ & $5.00(\textrm{N/A})$ & $200.00(\textrm{N/A})$ & $120.00(\textrm{N/A})$ & & $10.00(\textrm{N/A})$ & $10.00(\textrm{N/A})$ & $20.00(\textrm{N/A})$ & $1.00(\textrm{N/A})$ & $5.02(\textrm{N/A})$ & & $0.93$ & $7.64\!\times\!10^{-4}$ & \\
17 & $8.00(\textrm{N/A})$ & $5.00(\textrm{N/A})$ & $200.00(\textrm{N/A})$ & $120.00(\textrm{N/A})$ & & $9.99(\textrm{N/A})$ & $10.02(\textrm{N/A})$ & $20.00(\textrm{N/A})$ & $1.01(\textrm{N/A})$ & $15.00(\textrm{N/A})$ & & $0.81$ & $9.25\!\times\!10^{-4}$ & \\
18 & $8.04(\textrm{N/A})$ & $5.00(\textrm{N/A})$ & $200.00(\textrm{N/A})$ & $119.86(\textrm{N/A})$ & & $9.96(\textrm{N/A})$ & $9.95(\textrm{N/A})$ & $20.02(\textrm{N/A})$ & $0.98(\textrm{N/A})$ & $29.97(\textrm{N/A})$ & & $0.96$ & $3.89\!\times\!10^{-3}$ & \\
19 & $8.00(\textrm{N/A})$ & $5.00(\textrm{N/A})$ & $200.00(\textrm{N/A})$ & $120.03(\textrm{N/A})$ & & $10.01(\textrm{N/A})$ & $9.97(\textrm{N/A})$ & $20.00(\textrm{N/A})$ & $1.48(\textrm{N/A})$ & $5.03(\textrm{N/A})$ & & $1.06$ & $2.64\!\times\!10^{-3}$ & \\
20 & $7.98(\textrm{N/A})$ & $5.00(\textrm{N/A})$ & $200.00(\textrm{N/A})$ & $120.00(\textrm{N/A})$ & & $10.02(\textrm{N/A})$ & $10.07(\textrm{N/A})$ & $20.00(\textrm{N/A})$ & $1.51(\textrm{N/A})$ & $14.97(\textrm{N/A})$ & & $0.90$ & $2.32\!\times\!10^{-3}$ & \\
21 & $8.00(0.03)$ & $5.00(0.06)$ & $199.99(19.61)$ & $120.03(0.07)$ & & $9.99(0.20)$ & $9.99(0.20)$ & $20.00(0.17)$ & $1.50(0.45)$ & $29.97(0.59)$ & & $1.00$ & $7.71\!\times\!10^{-4}$ & \\
22 & $8.00(\textrm{N/A})$ & $5.00(\textrm{N/A})$ & $200.01(\textrm{N/A})$ & $119.99(\textrm{N/A})$ & & $10.00(\textrm{N/A})$ & $9.99(\textrm{N/A})$ & $20.00(\textrm{N/A})$ & $2.00(\textrm{N/A})$ & $5.01(\textrm{N/A})$ & & $0.93$ & $4.20\!\times\!10^{-4}$ & \\
23 & $8.01(\textrm{N/A})$ & $5.00(\textrm{N/A})$ & $200.00(\textrm{N/A})$ & $119.98(\textrm{N/A})$ & & $10.00(\textrm{N/A})$ & $9.99(\textrm{N/A})$ & $20.00(\textrm{N/A})$ & $2.01(\textrm{N/A})$ & $15.00(\textrm{N/A})$ & & $1.05$ & $6.85\!\times\!10^{-4}$ & \\
24 & $8.00(10^{-3})$ & $5.00(0.01)$ & $200.00(9.02)$ & $119.99(0.04)$ & & $10.00(0.02)$ & $9.99(0.02)$ & $20.00(0.02)$ & $2.00(0.25)$ & $30.00(0.06)$ & & $0.90$ & $2.49\!\times\!10^{-4}$ & \\
25 & $8.00(\textrm{N/A})$ & $5.01(\textrm{N/A})$ & $200.08(\textrm{N/A})$ & $119.92(\textrm{N/A})$ & & $10.02(\textrm{N/A})$ & $10.03(\textrm{N/A})$ & $20.00(\textrm{N/A})$ & $2.51(\textrm{N/A})$ & $4.98(\textrm{N/A})$ & & $1.03$ & $1.98\!\times\!10^{-3}$ & \\
26 & $8.00(\textrm{N/A})$ & $5.00(\textrm{N/A})$ & $200.07(\textrm{N/A})$ & $120.00(\textrm{N/A})$ & & $10.00(\textrm{N/A})$ & $10.03(\textrm{N/A})$ & $20.00(\textrm{N/A})$ & $2.52(\textrm{N/A})$ & $14.93(\textrm{N/A})$ & & $0.82$ & $1.80\!\times\!10^{-3}$ & \\
27 & $8.00(\textrm{N/A})$ & $5.00(\textrm{N/A})$ & $200.01(\textrm{N/A})$ & $120.00(\textrm{N/A})$ & & $10.00(\textrm{N/A})$ & $10.00(\textrm{N/A})$ & $20.00(\textrm{N/A})$ & $2.50(\textrm{N/A})$ & $30.00(\textrm{N/A})$ & & $1.04$ & $7.68\!\times\!10^{-5}$ & Smallest $\left<\delta_\lambda\right>$ \\
28 & $8.05(\textrm{N/A})$ & $5.01(\textrm{N/A})$ & $199.99(\textrm{N/A})$ & $120.01(\textrm{N/A})$ & & $9.93(\textrm{N/A})$ & $9.90(\textrm{N/A})$ & $20.01(\textrm{N/A})$ & $3.01(\textrm{N/A})$ & $4.89(\textrm{N/A})$ & & $0.94$ & $5.76\!\times\!10^{-3}$ & \\
29 & $8.01(\textrm{N/A})$ & $5.00(\textrm{N/A})$ & $200.01(\textrm{N/A})$ & $119.94(\textrm{N/A})$ & & $9.99(\textrm{N/A})$ & $9.99(\textrm{N/A})$ & $20.00(\textrm{N/A})$ & $3.01(\textrm{N/A})$ & $15.00(\textrm{N/A})$ & & $0.94$ & $7.12\!\times\!10^{-4}$ & \\
30 & $8.00(\textrm{N/A})$ & $4.99(\textrm{N/A})$ & $199.99(\textrm{N/A})$ & $119.97(\textrm{N/A})$ & & $9.96(\textrm{N/A})$ & $9.96(\textrm{N/A})$ & $20.01(\textrm{N/A})$ & $3.01(\textrm{N/A})$ & $30.20(\textrm{N/A})$ & & $0.94$ & $2.26\!\times\!10^{-3}$ & \\
\\[-1em]
\hline
\\[-1em]
31 & $8.00(\textrm{N/A})$ & $5.00(\textrm{N/A})$ & $200.02(\textrm{N/A})$ & $120.03(\textrm{N/A})$ & & $30.00(\textrm{N/A})$ & $10.05(\textrm{N/A})$ & $20.00(\textrm{N/A})$ & $1.06(\textrm{N/A})$ & $5.01(\textrm{N/A})$ & & $1.02$ & $7.18\!\times\!10^{-3}$ & \\
32 & $8.00(0.01)$ & $4.99(0.01)$ & $200.01(133.94)$ & $119.94(0.08)$ & & $30.02(0.20)$ & $10.02(0.12)$ & $20.00(0.03)$ & $1.04(2.71)$ & $14.97(0.33)$ & & $1.13$ & $5.77\!\times\!10^{-3}$ & \\ 
33 & $8.00(\textrm{N/A})$ & $5.00(\textrm{N/A})$ & $200.02(\textrm{N/A})$ & $119.90(\textrm{N/A})$ & & $30.03(\textrm{N/A})$ & $10.04(\textrm{N/A})$ & $19.99(\textrm{N/A})$ & $1.13(\textrm{N/A})$ & $30.01(\textrm{N/A})$ & & $1.02$ & $1.54\!\times\!10^{-2}$ & \\
34 & $7.99(0.01)$ & $4.99(0.03)$ & $199.99(47.42)$ & $119.95(0.08)$ & & $30.02(0.41)$ & $10.00(0.13)$ & $20.00(0.15)$ & $1.53(1.17)$ & $5.16(0.94)$ & & $1.19$ & $5.96\!\times\!10^{-3}$ & \\
35 & $8.00(\textrm{N/A})$ & $5.00(\textrm{N/A})$ & $200.02(\textrm{N/A})$ & $119.97(\textrm{N/A})$ & & $30.02(\textrm{N/A})$ & $10.04(\textrm{N/A})$ & $19.99(\textrm{N/A})$ & $1.59(\textrm{N/A})$ & $14.98(\textrm{N/A})$ & & $1.02$ & $7.69\!\times\!10^{-3}$ & \\
36 & $8.00(\textrm{N/A})$ & $5.00(\textrm{N/A})$ & $200.00(\textrm{N/A})$ & $120.00(\textrm{N/A})$ & & $30.01(\textrm{N/A})$ & $10.02(\textrm{N/A})$ & $20.00(\textrm{N/A})$ & $1.55(\textrm{N/A})$ & $29.99(\textrm{N/A})$ & & $1.02$ & $4.22\!\times\!10^{-3}$ & \\
37 & $8.06(0.08)$ & $5.04(0.11)$ & $200.01(75.29)$ & $120.07(0.07)$ & & $29.76(2.83)$ & $9.95(0.95)$ & $19.99(0.92)$ & $2.10(2.24)$ & $5.10(6.68)$ & & $1.06$ & $1.13\!\times\!10^{-2}$ & \\
38 & $7.98(\textrm{N/A})$ & $5.01(\textrm{N/A})$ & $200.00(\textrm{N/A})$ & $119.97(\textrm{N/A})$ & & $30.12(\textrm{N/A})$ & $10.00(\textrm{N/A})$ & $19.99(\textrm{N/A})$ & $1.92(\textrm{N/A})$ & $15.07(\textrm{N/A})$ & & $0.94$ & $5.87\!\times\!10^{-3}$ & \\
39 & $8.00(\textrm{N/A})$ & $5.00(\textrm{N/A})$ & $200.00(\textrm{N/A})$ & $120.00(\textrm{N/A})$ & & $30.00(\textrm{N/A})$ & $10.00(\textrm{N/A})$ & $20.00(\textrm{N/A})$ & $2.00(\textrm{N/A})$ & $30.00(\textrm{N/A})$ & & $1.03$ & $2.25\!\times\!10^{-4}$ & 2nd smallest $\left<\delta_\lambda\right>$ \\
40 & $7.99(0.15)$ & $5.00(0.20)$ & $200.01(37.97)$ & $119.94(0.06)$ & & $30.07(6.30)$ & $9.99(2.09)$ & $20.01(2.05)$ & $2.44(1.18)$ & $4.92(13.44)$ & & $1.09$ & $5.10\!\times\!10^{-3}$ & \\
41 & $8.00(\textrm{N/A})$ & $5.00(\textrm{N/A})$ & $200.00(\textrm{N/A})$ & $120.02(\textrm{N/A})$ & & $29.97(\textrm{N/A})$ & $10.01(\textrm{N/A})$ & $20.00(\textrm{N/A})$ & $2.54(\textrm{N/A})$ & $14.99(\textrm{N/A})$ & & $1.11$ & $2.00\!\times\!10^{-3}$ & \\
42 & $7.96(\textrm{N/A})$ & $5.00(\textrm{N/A})$ & $199.98(\textrm{N/A})$ & $119.91(\textrm{N/A})$ & & $30.27(\textrm{N/A})$ & $10.06(\textrm{N/A})$ & $20.00(\textrm{N/A})$ & $2.45(\textrm{N/A})$ & $29.99(\textrm{N/A})$ & & $0.90$ & $4.74\!\times\!10^{-3}$ \\
43 & $8.00(\textrm{N/A})$ & $5.00(\textrm{N/A})$ & $200.00(\textrm{N/A})$ & $119.99(\textrm{N/A})$ & & $30.00(\textrm{N/A})$ & $9.98(\textrm{N/A})$ & $20.00(\textrm{N/A})$ & $2.99(\textrm{N/A})$ & $5.01(\textrm{N/A})$ & & $0.99$ & $8.90\!\times\!10^{-4}$ & \\
44 & $7.98(\textrm{N/A})$ & $4.99(\textrm{N/A})$ & $200.00(\textrm{N/A})$ & $119.96(\textrm{N/A})$ & & $30.06(\textrm{N/A})$ & $9.99(\textrm{N/A})$ & $20.00(\textrm{N/A})$ & $2.89(\textrm{N/A})$ & $15.15(\textrm{N/A})$ & & $0.98$ & $6.14\!\times\!10^{-3}$ & \\
45 & $8.00(\textrm{N/A})$ & $5.00(\textrm{N/A})$ & $199.99(\textrm{N/A})$ & $120.05(\textrm{N/A})$ & & $29.98(\textrm{N/A})$ & $9.99(\textrm{N/A})$ & $20.00(\textrm{N/A})$ & $2.95(\textrm{N/A})$ & $30.00(\textrm{N/A})$ & & $1.06$ & $1.96\!\times\!10^{-3}$ &  \\
\\[-1em]
\hline
\end{tabular}
\end{table*}

We perform the fitting with the prior range tabulated in Table~\ref{tab:prior} and summarize the result obtained using the method in a Python package, \textsc{lmfit}~\citep{lmfit}, in Table~\ref{tab:nm_fitting} for simulated samples prepared with the combination of different values of selected parameters $q$, $s$, $e$, and $\iota$ including the samples studied in the previous section. Here, we test following parameter values, $q=\{0.3, 1.0\}$, $s=\{10~\textrm{AU}, 20~\textrm{AU}\}$, $e=\{1.0,1.5,2.0,2.5,3.0\}$, and $\iota=\{5^\circ, 15^\circ, 30^\circ\}$. 

\begin{figure*}
\centering
    \includegraphics[width=0.322\linewidth]{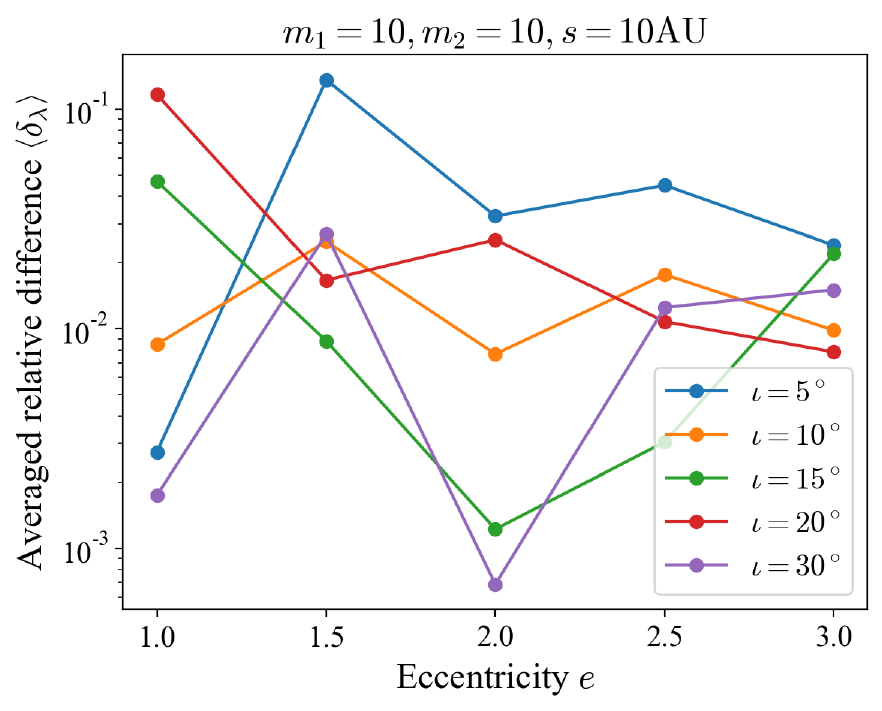}
    \includegraphics[width=0.322\linewidth]{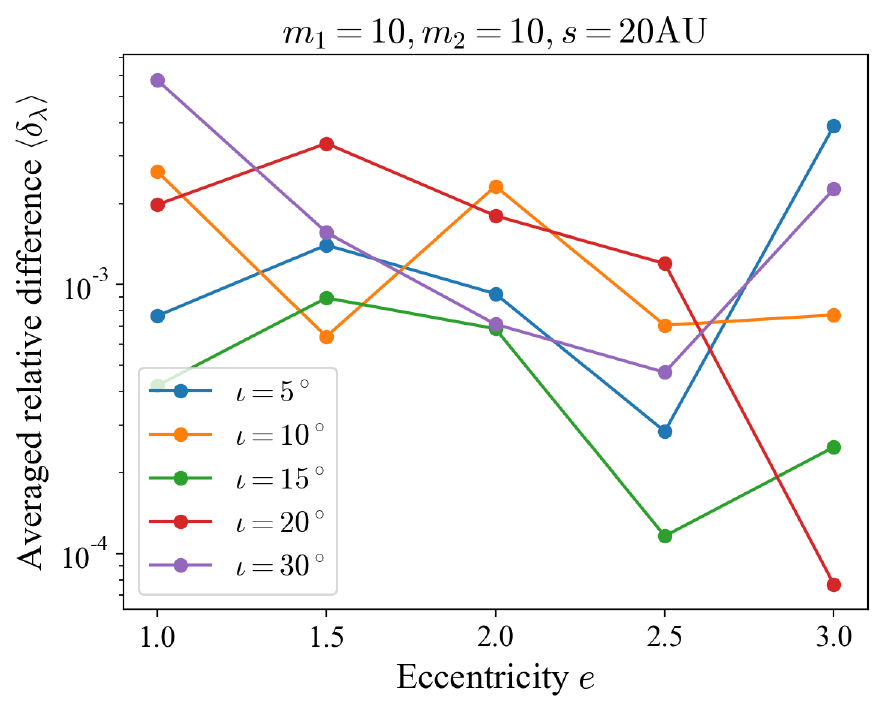}
    \includegraphics[width=0.322\linewidth]{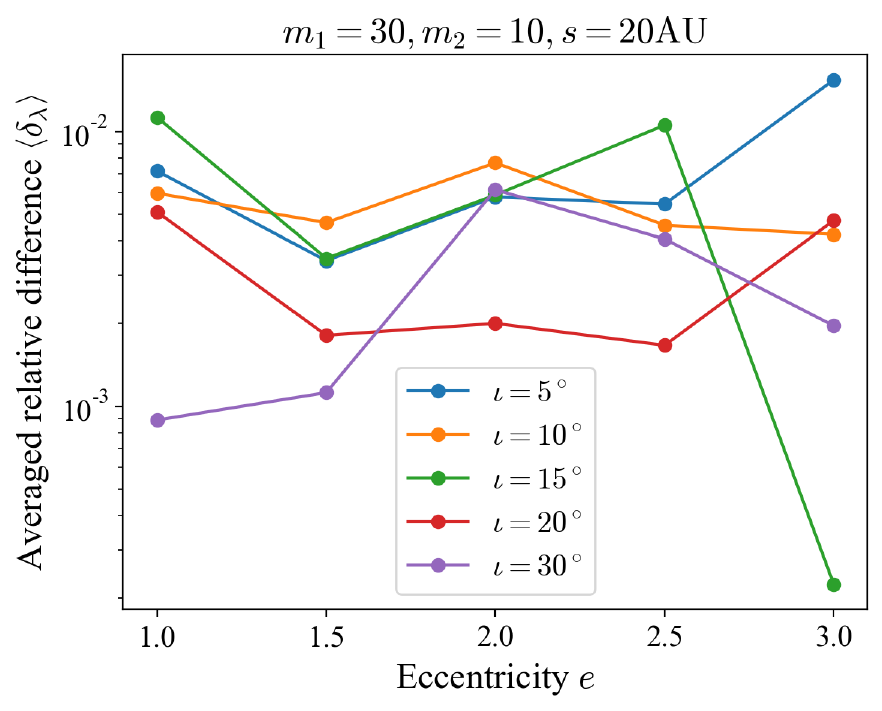}
    \caption{True value of eccentricity $e$ versus averaged relative difference $\left<\delta_\lambda\right>$ of Case I (left), II (middle), and III (right). We collect the result of five different values of $\iota$, i.e., $\iota=5^\circ$, $10^\circ$, $15^\circ$, $20^\circ$, and $30^\circ$, into one figure of each case for comparison. One can see that there is no correlation between $e$ and $\left<\delta_\lambda\right>$. Also, this result shows that the value of $\iota$ is unrelated to $\left<\delta_\lambda\right>$ with given $e$. \label{fig:delta_r_vs_e}}
\end{figure*}

\begin{figure*}
\centering
    \subfigure[Sample 27 (Smallest $\left<\delta_\lambda\right>$)]{
        \includegraphics[width=0.48\linewidth]{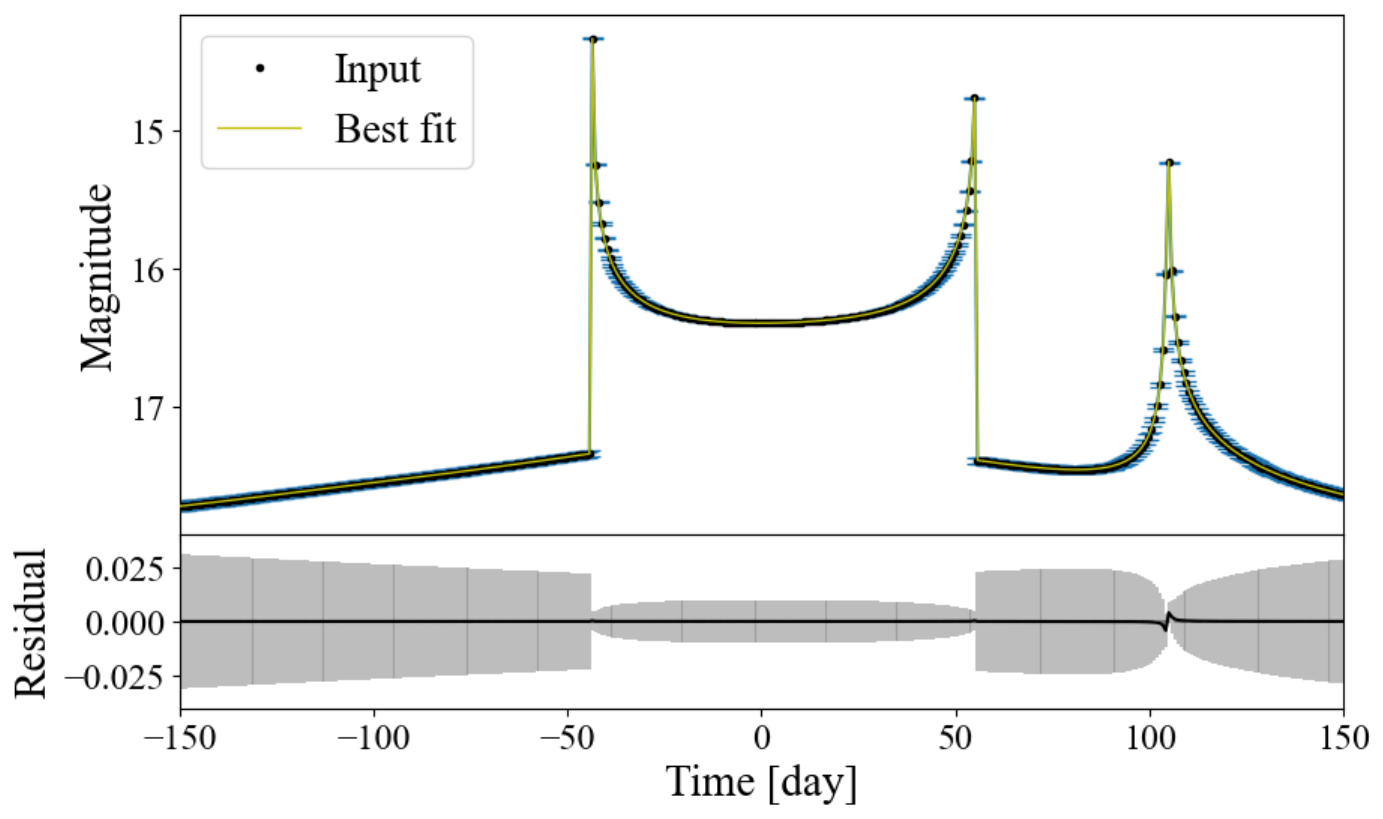}
    }
    \subfigure[Sample 10 (Largest $\left<\delta_\lambda\right>$)]{
        \includegraphics[width=0.48\linewidth]{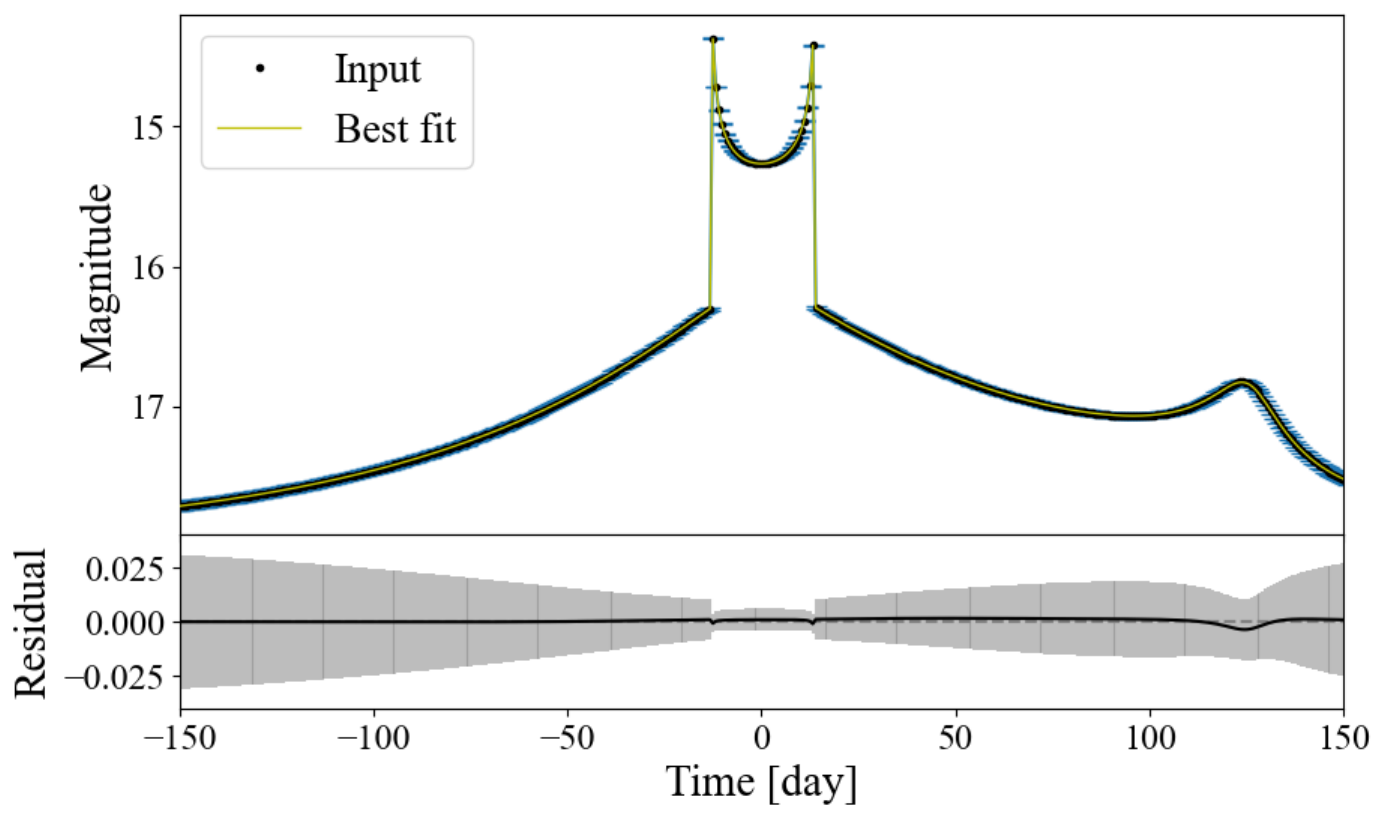}
    }
    \caption{Input and best fit light curves of Sample 27 (left) and 10 (right) corresponding to the smallest and the largest $\left<\delta_\lambda\right>$, respectively. \label{fig:bestfit_lc}}
\end{figure*}

As tabulated in Table~\ref{tab:nm_fitting}, in order to measure the goodness-of-fitting, we compute the reduced $\chi^2$, i.e., $\chi^2_r$, which is defined as
\begin{equation}
\chi^2_r = \sum^N_i \rho_i^2 / (N - N_\textrm{vars})~, \label{eq:chi_r}
\end{equation}
where $\rho_i$ is the residual of magnitude subtracting the inferred magnitude $d({\Lambda})_{i}$ from the true magnitude $d_{T,i}$ and scaled by the data uncertainty $\sigma_i$, i.e., 
\begin{equation}
\rho_i = \frac{d_{T,i} - d({\Lambda})_{i}}{\sigma_i}
\end{equation}
at each observation point $i$. In this work, we adopt the same form of $\sigma_i$ used in \cite{Ma:2021ucq} which is given by 
\begin{equation}
\sigma_i = \sqrt{\sigma_\textrm{sys}^2 + \sigma_0^2 10^{0.8(d_{T,i} - d_0)}},
\end{equation}
where $\sigma_\textrm{sys}=0.004$ is the systematic floor and $\sigma_0=0.04$ is the uncertainty at the reference magnitude $d_0$ set as 18. $N$ and $N_\textrm{vars}$ denote the number of observation data points and the number of variables that are fixed as $400$ and $9$ in this work, respectively. Note that we take $\sum_i\rho_i$ as the cost function of the NM method for seeking the best fitting parameters. Along with $\chi_r^2$, for the figure-of-merit of the fitting-based parameter estimation, we calculate the averaged relative difference $\left<\delta_\lambda\right>$ as follows
\begin{equation}
\left<\delta_\lambda\right> = \sum_\lambda \delta_\lambda / N_\textrm{vars}~, \label{eq:delta_r_mean}
\end{equation}
where $\delta_\lambda$ is the relative difference between the true value $\lambda_T$ and inferred value $\lambda_I$ of each parameter $\lambda$, i.e., $\delta_\lambda = |\lambda_T - \lambda_I|/\lambda_T$, and summarize the value for each sample in Table~\ref{tab:nm_fitting}.

Looking at Table~\ref{tab:nm_fitting}, we can see that the 9 parameters of binary microlensing are mostly well recovered by the NM-based fitting. Also, the goodness-of-fit, represented by the value of $\chi^2_r$, of all tested samples results in around $1$ which means the extent of residuals is in accord with given degrees of freedom, i.e., $(N-N_\textrm{var})$, in the fitting. When we compare the range of $\left<\delta_\lambda\right>$ of samples in each case, the range of $\left<\delta_\lambda\right>$ of Case I, resulted in $\sim\!10^{-3}$--$10^{-2}$, are relatively larger than the ranges of Case II and III, $\sim\!10^{-4}$--$10^{-3}$ and $\sim\!10^{-4}$--$10^{-2}$, respectively. This result tells us that relatively smaller/narrower caustics of shorter $s$ such as $s=10~\textrm{AU}$ hinders precise estimations than longer $s$ such as $s=20~\textrm{AU}$.

Next, we examine the relation between the value of $e$ and $\left<\delta_\lambda\right>$ because our main interest is whether the precise parameter estimation is possible or not for a BH binary lens of high eccentric orbit such as $e\geq1$. In order to examine the correlation, we draw the plot of $e$ versus $\left<\delta_\lambda\right>$ in Fig.~\ref{fig:delta_r_vs_e}. Also, for comparison, we collect the result of five different values of $\iota$, i.e., $\iota=\{5^\circ$, $10^\circ$, $15^\circ$, $20^\circ$, $30^\circ\}$, into the figure of each case. One can see from the subfigures of Fig.~\ref{fig:delta_r_vs_e} that there is no $e$-dependency in $\left<\delta_\lambda\right>$. In other words, it is possible to obtain low $\left<\delta_\lambda\right>$ even for a BH binary of higher $e$. The lesson we can learn from this result is that there is no obvious hindrance in the parameter estimation of a BH binary of $e\geq1$ from binary microlensing light curves, if a well-prepared model is applied to the fitting. Beside, we can see that $\left<\delta_\lambda\right>$ is also independent from the value of $\iota$. Hence, the NM-based fitting with the considered binary lens model allows us to recognise the $\iota$-dependent changes in $s$ and $e$.

From this fitting-based study, it turns out that the values of $\left<\delta_\lambda\right>$ of Sample 27 and 39 are the smallest and second smallest $\left<\delta_\lambda\right>$ while Sample 10 and 7 are the largest and second largest $\left<\delta_\lambda\right>$, respectively, amongst all tested samples. In specific, we can see that the estimated value of $\iota$ of Sample 10 and 7 are more different from the true value, $+5.00^\circ$ and $+1.89^\circ$, respectively, than other samples which mostly resulted in less than  $\pm1^\circ$ deviations from the true $\iota$. Despite of it, the difference between the true value and estimated value of other parameters are less than $10\%$ of the true value. Indeed, when we plot the light curve with the best fit parameters, as presented in Fig.~\ref{fig:bestfit_lc} for Sample 27 and 10, we see that the mismatch between the input and best fit light curves mainly happens at peaks/bumps and the best fit light curves can more or less perfectly reconstruct input light curves.

On the other hand, the NM fitting for most samples of Case II and III failed in estimating uncertainties of all 9 parameters. This is because, for example, (i) if a variable has no practical contribution on the fit or (ii) if the fit estimate the value of a variable being near the maximum or minimum of given bounds, the variable likely causes the covariance matrix singular that making standard errors impossible to estimate~\citep{lmfit}. In reality, the uncertainty of estimated parameter is quite important because it not only gives us confidence in the estimation but also it is helpful to conduct follow-up analyzes
with the given range of uncertainty. Therefore, we need an alternative of parameter estimation that allows us to estimate uncertainties.

\begin{table*}
\small
\caption{True value $\lambda_T$ and inferred value $\lambda_I$ for the 9 model parameters of Sample 27 and 10 corresponding to the smallest and the largest $\left<\delta_\lambda\right>$ values in Table~\ref{tab:nm_fitting}. For comparison, we tabulate $\lambda_{I,\textrm{NM}}$ and $\delta_{\lambda,\textrm{NM}}$ together with $\lambda_{I,\textrm{MCMC}}$ and $\delta_{\lambda,\textrm{MCMC}}$. The uncertainty given for $\lambda_I$ represents $1\sigma$ error, except $\lambda_{I,\textrm{NM}}$ that were failed in the uncertainty estimation from the NM fitting. We also summarize $\delta_\lambda$ of each parameter calculated with the true value and the median value.}
\label{tab:mcmc_result}
\centering
\begin{tabular}{@{} c c c c c c C{0.01cm} c c c c c @{}}
\hline
\multirow{2}{*}{$\lambda$ (unit)} & \multicolumn{5}{c}{Sample 27} & & \multicolumn{5}{c}{Sample 10} \\
\cline{2-6} \cline{8-12}\\[-1em]
& $\lambda_T$ & $\lambda_{I,\textrm{NM}}$ & $\delta_{\lambda,\textrm{NM}}$ & $\lambda_{I,\textrm{MCMC}}$ & $\delta_{\lambda,\textrm{MCMC}}$ & & $\lambda_T$ & $\lambda_{I,\textrm{NM}}$ & $\delta_{\lambda,\textrm{NM}}$ & $\lambda_{I,\textrm{MCMC}}$ & $\delta_{\lambda,\textrm{MCMC}}$ \\
\\[-1em]
\hline
\\[-1em]
$D_S$ (kpc) & $8$ & $8.00\!\pm\!\textrm{N/A}$ & $2.42\!\times\!10^{-5}$ & $8.01_{-0.29}^{+0.57}$ & $7.71\!\times\!10^{-4}$ & & $8$ & $7.95\!\pm\!0.34$ & $6.18\!\times\!10^{-3}$ & $8.66_{-1.48}^{+1.62}$ & $8.28\!\times\!10^{-2}$ \\
\\[-1em]
$D_L$ (kpc) & $5$ & $5.00\!\pm\!\textrm{N/A}$ & $6.09\!\times\!10^{-5}$ & $5.00_{-0.19}^{+0.28}$ & $4.90\!\times\!10^{-5}$ & & $5$ & $5.09\!\pm\!0.78$ & $1.78\!\times\!10^{-2}$ & $5.45_{-0.88}^{+0.85}$ & $8.95\!\times\!10^{-2}$ \\
\\[-1em]
$v_r$ (km/s) & $200$ & $200.01\!\pm\!\textrm{N/A}$ & $3.32\!\times\!10^{-5}$ & $201.94_{-7.33}^{+5.44}$ & $9.72\!\times\!10^{-3}$ & & $200$ & $199.93\!\pm\!29.25$ & $3.64\!\times\!10^{-4}$ & $201.00_{-5.49}^{+5.38}$ & $4.99\!\times\!10^{-3}$ \\
\\[-1em]
$\theta$ ($^\circ$) & $120$ & $120.00\!\pm\!\textrm{N/A}$ & $1.68\!\times\!10^{-5}$ & $120.00_{-0.02}^{+0.03}$ & $6.07\!\times\!10^{-6}$ & & $120$ & $119.89\!\pm\!0.08$ & $9.07\!\times\!10^{-4}$ & $119.93_{-0.08}^{+0.08}$ & $5.92\!\times\!10^{-4}$ \\
\\[-1em]
$m_1$ ($M_\odot$) & $10$ & $10.00\!\pm\!\textrm{N/A}$ & $5.07\!\times\!10^{-5}$ & $10.16_{-0.46}^{+0.40}$ & $1.58\!\times\!10^{-2}$ & & $10$ & $9.94\!\pm\!2.19$ & $5.78\!\times\!10^{-3}$ & $11.08_{-1.96}^{+2.00}$ & $1.08\!\times\!10^{-1}$ \\
\\[-1em]
$m_2$ ($M_\odot$) & $10$ & $10.00\!\pm\!\textrm{N/A}$ & $8.16\!\times\!10^{-5}$ & $10.16_{-0.46}^{+0.40}$ & $1.56\!\times\!10^{-2}$ & & $10$ & $9.94\!\pm\!2.19$ & $5.53\!\times\!10^{-3}$ & $11.09_{-1.95}^{+2.02}$ & $1.09\!\times\!10^{-1}$ \\
\\[-1em]
$s$ (AU) & $20$ & $20.00\!\pm\!\textrm{N/A}$ & $7.95\!\times\!10^{-5}$ & $20.30_{-0.99}^{+1.16}$ & $2.79\!\times\!10^{-2}$ & & $10$ & $9.97\!\pm\!0.92$ & $3.44\!\times\!10^{-3}$ & $10.83_{-1.73}^{+2.38}$ & $8.34\!\times\!10^{-2}$ \\
\\[-1em]
$e$ (--) & $2.5$ & $2.50\!\pm\!\textrm{N/A}$ & $3.20\!\times\!10^{-4}$ & $2.57_{-0.25}^{+0.17}$ & $5.44\!\times\!10^{-3}$ & & $2.5$ & $2.48\!\pm\!0.49$ & $9.24\!\times\!10^{-3}$ & $2.50_{-0.11}^{+0.10}$ & $1.38\!\times\!10^{-3}$ \\
\\[-1em]
$\iota$ ($^\circ$) & $30$ & $30.00\!\pm\!\textrm{N/A}$ & $2.46\!\times\!10^{-5}$ & $30.16_{-0.83}^{+0.93}$ & $7.16\!\times\!10^{-4}$ & & $5$ & $10.00\!\pm\!24.39$ & $1.00\!\times\!10^{0}$ & $4.99_{-4.78}^{+5.39}$ & $1.18\!\times\!10^{-3}$ \\
\hline
\end{tabular}
\end{table*}

\begin{figure*}
\centering
    \includegraphics[width=.9\linewidth]{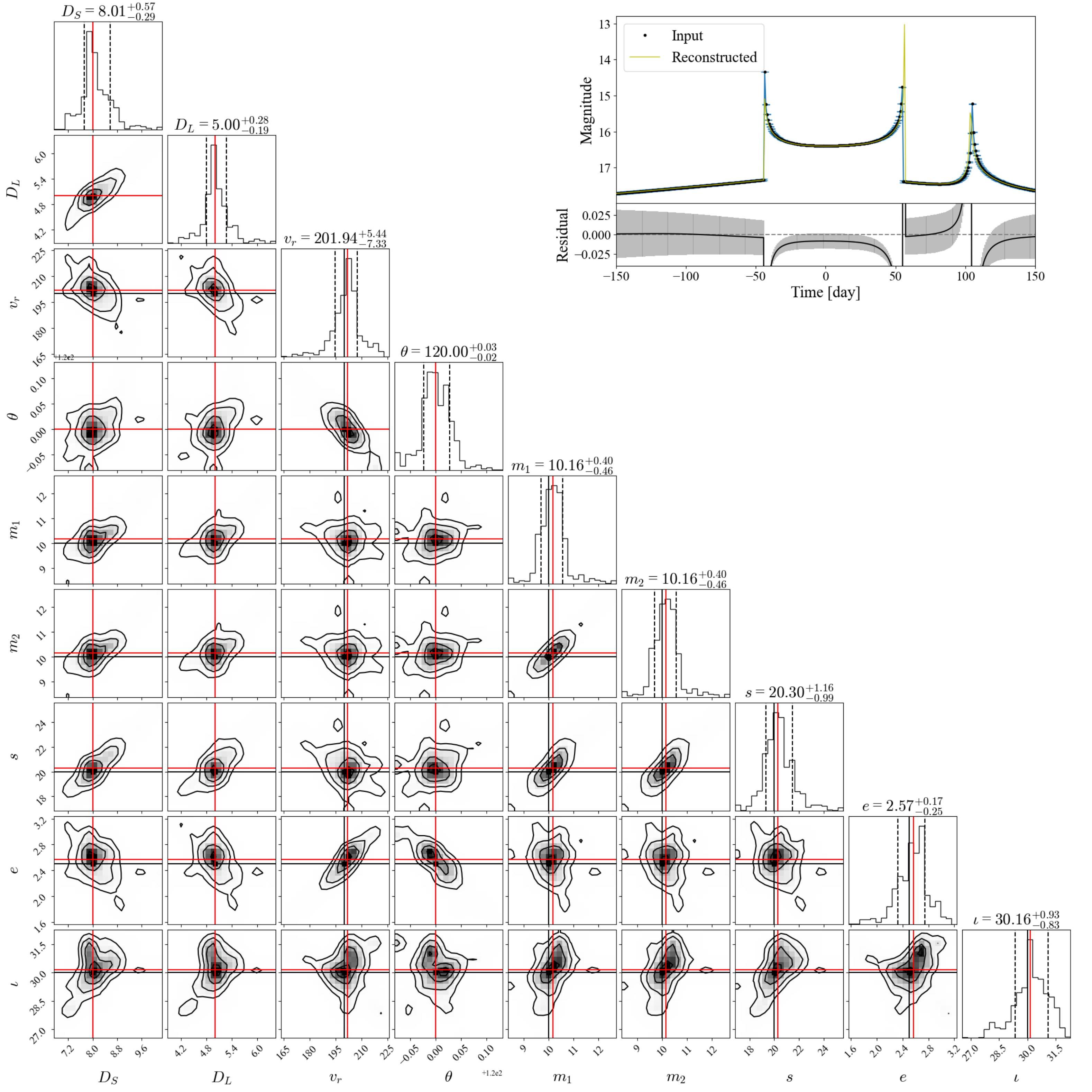} 
    \caption{Results of MCMC-based parameter estimation for Sample 27. The corner plot represents posterior distributions of 9 parameters inferred by the MCMC-based analysis. We mark the true value as black solid line and the median value as red solid line with the $1\sigma$ uncertainty depicted as black dashed line. The upper-right inset figure shows the reconstructed light curve using the median values of 9 parameters and the residuals after the subtraction of reconstructed light curve from the input. \label{fig:corner_lc_sample_27}}
\end{figure*}

\begin{figure*}
\centering
    \includegraphics[width=.9\linewidth]{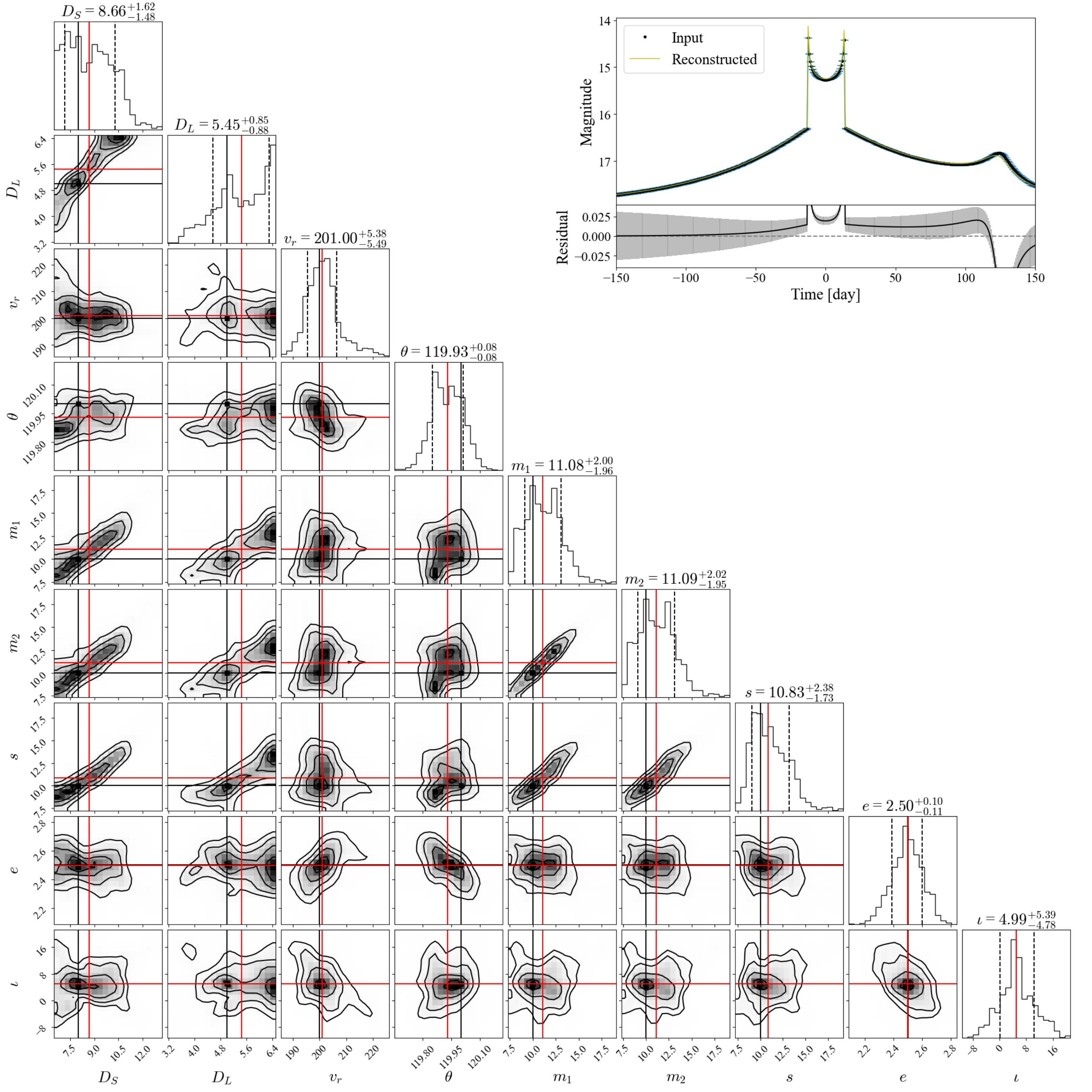}
    \caption{Results of MCMC-based parameter estimation for Sample 10. Similar to Fig.~\ref{fig:corner_lc_sample_27}, we present corner plot and reconstructed light curve.  \label{fig:corner_lc_sample_10}}
\end{figure*}

In this work, we consider the Markov chain Monte Carlo (MCMC)-based Bayesian inference~\citep{mcmc} as an alternative of parameter estimation. For the Bayesian inference, we conduct sampling of the posterior probability density of parameters of interest using the ensemble sampler implemented in the \textsc{emcee}~\citep{emcee} package. For consistency, we take the same prior range used in the NM fitting. Finding the global minimum is tested with the log-likelihood function of a $\Lambda$-dependent light curve data, given the input data $d$, which is defined as
\begin{equation}
\ln P(d|{\Lambda}) = -\frac{1}{2}\sum_i \left[ \frac{(d({\Lambda})_i - d_i)^2}{\sigma_i^2} + \ln 2\pi \sigma_i^2 \right], 
\end{equation}
where $d({\Lambda})$ is the light curve drawn by the model with a given parameter set ${\Lambda}$ and $\sigma_i$ is the same data uncertainty taken for the calculation of $\rho_i$ in the NM fitting. Then, we let each MCMC process take 20 walkers and each walker draws 5,000 samples from the posterior distribution.

In Table~\ref{tab:mcmc_result}, we summarize the result of the MCMC-based parameter estimation for Sample 27 that was resulted in the smallest $\left<\delta_\lambda\right>$ but the uncertainty estimation was failed in the NM fitting. Also, we examine Sample 10 resulted in the largest $\left<\delta_\lambda\right>$ although its uncertainty estimation is done even with the NM fitting. One can see from $\lambda_{I,\textrm{MCMC}}$ of Sample 27 that the uncertainty estimation for each inferred value is available with the MCMC-based parameter estimation as desired. On top of that, the estimated uncertainty successfully covers each $\lambda_T$ although the median value itself is slightly deviated from $\lambda_T$. However, when we compare $\delta_{\lambda,\textrm{MCMC}}$ with $\delta_{\lambda,\textrm{NM}}$ for both Sample 27 and 10, it is shown that the deviation of $\lambda_{I,\textrm{MCMC}}$ from $\lambda_T$ is mostly bigger than the deviation of $\lambda_{I,\textrm{NM}}$ from the same $\lambda_T$. In consequence, for the two samples, $\left<\delta_\lambda\right>$ is calculated as $1.00\times10^{-2}$ and $5.35\times10^{-2}$, respectively, that were $7.68\times10^{-5}$ and $1.17\times10^{-1}$ from the NM fitting. Despite this, the uncertainty estimated from the MCMC-based parameter estimation results in better precision on several parameters, e.g., $v_r$, $m_1$, $m_2$, $e$, and $\iota$, than the NM-based parameter estimation. In specific, the parameter estimation of $e$ and $\iota$ is improved by the MCMC-based parameter estimation in terms of not only the precision but also the accuracy.

\begin{table}
\addtolength{\tabcolsep}{-0.25em}
\small
\caption{Inferred parameter value $\lambda_I$ for Sample 10, obtained from the initial NM-based fitting ($\lambda_{I,\textrm{NM}}$) and from refitting with NM method based on the $1\sigma$ error bar of MCMC result ($\lambda_{I,\textrm{NM}_\textrm{refit}}$) for the 9 model parameters of Sample 10. \label{tab:NM_rerun_sample_10}}
\centering
\begin{tabular}{@{} c c c c c @{}}
\hline
\\[-1em]
Parameter ($\lambda_T$) & $\lambda_{I,\textrm{NM}}$ & $\delta_{\lambda,\textrm{NM}}$ & $\lambda_{I,\textrm{NM}_\textrm{refit}}$ & $\delta_{\lambda,\textrm{NM}_\textrm{refit}}$ \\
\\[-1em]
\hline
\\[-1em]
$D_S$ ($8$ kpc) & $7.95\!\pm\!0.34$ & $6.18\!\times\!10^{-3}$ & $7.99\!\pm\!\textrm{N/A}$ & $6.73\!\times\!10^{-4}$ \\
\\[-1em]
$D_L$ ($5$ kpc) & $5.09\!\pm\!0.78$ & $1.78\!\times\!10^{-2}$ & $4.89\!\pm\!\textrm{N/A}$ & $2.15\!\times\!10^{-2}$ \\
\\[-1em]
$v_r$ ($200$ km/s) & $199.93\!\pm\!29.25$ & $3.64\!\times\!10^{-4}$ & $200.01\!\pm\!\textrm{N/A}$ & $4.81\!\times\!10^{-5}$ \\
\\[-1em]
$\theta$ ($120^\circ$) & $119.89\!\pm\!0.08$ & $9.07\!\times\!10^{-4}$ & $119.93\!\pm\!\textrm{N/A}$ & $5.45\!\times\!10^{-4}$ \\
\\[-1em]
$m_1$ ($10~M_\odot$) & $9.94\!\pm\!2.19$ & $5.78\!\times\!10^{-3}$ & $10.02\!\pm\!\textrm{N/A}$ & $1.64\!\times\!10^{-3}$ \\
\\[-1em]
$m_2$ ($10~M_\odot$) & $9.94\!\pm\!2.19$ & $5.53\!\times\!10^{-3}$ & $10.03\!\pm\!\textrm{N/A}$ & $2.95\!\times\!10^{-3}$ \\
\\[-1em]
$s$ ($10$ AU) & $9.97\!\pm\!0.92$ & $3.44\!\times\!10^{-3}$ & $10.07\!\pm\!\textrm{N/A}$ & $7.36\!\times\!10^{-3}$ \\
\\[-1em]
$e$ (2.5) & $2.48\!\pm\!0.49$ & $9.24\!\times\!10^{-3}$ & $2.50\!\pm\!\textrm{N/A}$ & $1.59\!\times\!10^{-3}$ \\
\\[-1em]
$\iota$ ($5^\circ$) & $10.00\!\pm\!24.39$ & $1.00\!\times\!10^{0}$ & $5.17\!\pm\!\textrm{N/A}$ & $3.36\!\times\!10^{-2}$ \\
\\[-1em]
\hline
\end{tabular}
\end{table}

\begin{figure}
    \subfigure[Best fit light curve of Sample 10 from the initial NM]{
        \includegraphics[width=1.\linewidth]{lc_nm_m1_10_m2_10_sep_10_ecc_2.5_inc_5_host_Galaxy.pdf}
    }
    \subfigure[Best fit light curve of Sample 10 from NM$_\textrm{refit}$]{
        \includegraphics[width=1.\linewidth]{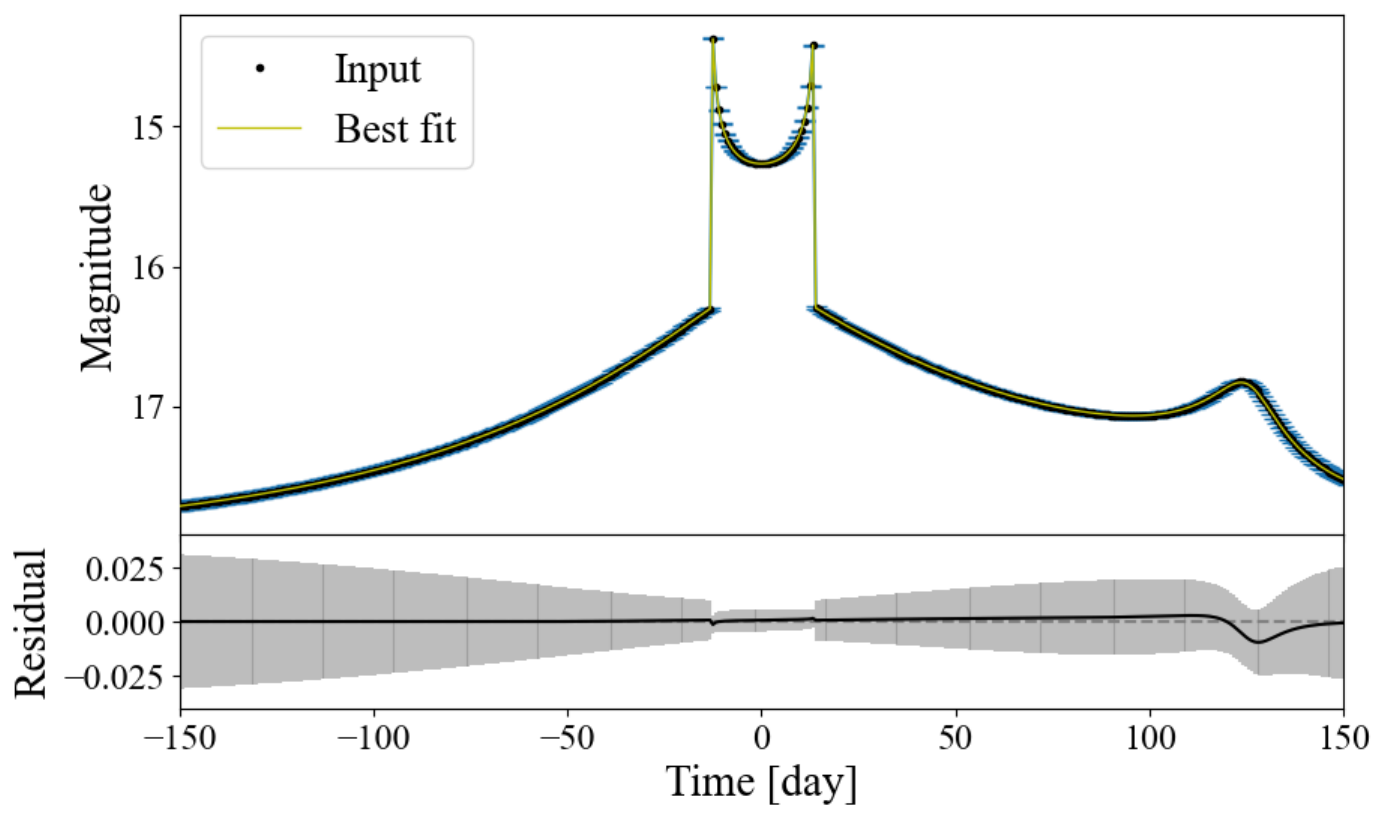}
    }
    \caption{Input and best fit light curves of Sample 10 resulted from the initial NM fitting (top) and the refitting (bottom), respectively. \label{fig:bestfit_lc_rerun}}
\end{figure}

We present the posterior distributions of the 9 selected parameters obtained from the MCMC-based parameter estimation. To this end, as shown in Figs.~\ref{fig:corner_lc_sample_27} (Sample 27) and \ref{fig:corner_lc_sample_10} (Sample 10), we draw the corner plot which consists of 1D histograms and 2D contour plots. We also depict the noisy input light curve and reconstructed light curve based on the median value of the posterior distribution of each parameter. In the 1D histogram and 2D contour plot, we mark the true value and the median of inferred values as the black-solid line and red-solid line, respectively. The black-dashed vertical lines indicates the lower and upper bounds of $1\sigma$ error for the median value. Looking at 1D histograms and 2D contour plots in Fig.~\ref{fig:corner_lc_sample_27}, we find that all posterior distributions peak around the true value and all true values place within the $1\sigma$ error bar. Similarly, we can see from Fig.~\ref{fig:corner_lc_sample_10} that the lower and upper bounds of $1\sigma$ error of all posterior distributions for Sample 10 contains true values successfully although there are some deviations between the true value and the inferred value that seen in Table~\ref{tab:mcmc_result}. However, when we look at the reconstructed light curves and the residuals presented in both Figs.~\ref{fig:corner_lc_sample_27} and \ref{fig:corner_lc_sample_10}, the mismatch between the input and reconstructed ones become much bigger than the best fit result shown in Fig.~\ref{fig:bestfit_lc} because of the more bigger difference in the inferred values of the MCMC-based result than that of the NM-based result.

To address the issue on bigger difference in the MCMC-based parameter estimation, we examine whether utilising the uncertainty information obtained from MCMC can be helpful in precise parameter estimation. As a demonstration, we provide the result from refitting Sample 10 with the NM method based on the $1\sigma$ error bar of the MCMC result in Table~\ref{tab:NM_rerun_sample_10} and Fig.~\ref{fig:bestfit_lc_rerun}. From Table~\ref{tab:NM_rerun_sample_10}, one can see improvements in $\delta_\lambda$ of each parameter, except $D_L$ and $s$, although $\delta_{D_L, \textrm{NM}_\textrm{refit}}$ and $\delta_{s, \textrm{NM}_\textrm{refit}}$ are still enhanced compared to $\delta_{D_L, \textrm{MCMC}}$ and $\delta_{s, \textrm{MCMC}}$, respectively. Also, the improvement in the inferred value of most parameters concludes to $\left<\delta_\lambda\right>$ too, from $\left<\delta_\lambda\right>_\textrm{NM}=1.17\times10^{-1}$ to $\left<\delta_\lambda\right>_{\textrm{NM}_\textrm{refit}}=7.76\times10^{-3}$, even though the refitting failed in uncertainty estimation for all parameters unlike the initial fitting. Therefore, we can confirm that utilising MCMC-based parameter estimation is indeed helpful in obtaining better estimation with a fitting-based parameter estimation. 

Despite of such improvement, when we compare the best fit light curves presented in Fig.~\ref{fig:bestfit_lc_rerun}, we observe that the residual around the right-hand-side bump is still larger than that of the initial NM fitting result. This result addresses not only that a relatively good fitting does not always guarantee the best parameter estimation but also that attaining the best result in both fitting and parameter estimation is quite nontrivial because of the degeneracy in the parameter-dependent light curve.


\section{Discussions}
\label{sec:discussions}


In this study, we have explored the characteristics of microlensing events caused by the stellar-mass BH binary, focusing on high eccentric lens orbits. We have compared the shapes of caustics and light curves in several selected cases, depending on the orbital properties of BH binary lens, and attempted to estimate the parameters that describe the lens and source system from simulated light curves. We have demonstrated parameter estimation by utilizing
two different methods---the Nelder-Mead fitting and a MCMC-based Bayesian inference---and confirmed that both approaches successfully recover the true parameter values within a reasonable uncertainty. 
The lesson we could learn from this study is
that 
microlensing by BHs of hyperbolic orbits
can even be distinguished from  
the microlensing by
elliptic BH binaries 
if adequate estimation methods are employed for given binary microlensing light curves.


However, we should note that we primarily examined the dynamical properties of the BH binary lens system, rather than exploring other parameters associated with the source and observer. In particular, as previously mentioned, microlensing parallax was disregarded due to its predicted small magnitude and the complexity of its evaluation for the BH binary lens. However, if we are able to appropriately account for the parallax effect through additional future work, more reliable parameter estimation will be achievable in actual observations.


To date, BH binaries have only been observed through GW detections. Due to the frequency range of the current GW detectors, the observed BH binaries have been limited to those with very small separations, just before merging. However, the BH system with an orbital size of about 10 AU, considered in this work, emits GWs at a frequency corresponding to the frequency band of Pulsar Timing Array (PTA). Moreover, if they are located in our Galaxy, the GW signals from these systems could be observed by the PTA using Square Kilometre Array. If the stellar-mass BH binaries are also observed through the microlensing effect by forthcoming telescopes such as Nancy Grace Roman Space Telescope, it would mark a significant milestone in multi-messenger astronomy, enriching our understanding of the BH population.


\begin{acknowledgements}
We thank Jeongcho Kim and Changsu Choi for fruitful discussions. This work is supported by the National Research Foundation of Korea (NRF) grants funded by the Ministry of Science and ICT (MSIT) of the Korea Government (NRF-2020R1C1C1005863, NRF-2021R1F1A1051269). The work of K.K. is partially supported by the Korea Astronomy and Space Science Institute under the R\&D program (Project No. 2024-1-810-02) supervised by the MSIT. K.K. also acknowledges that the computational work reported in this paper was partly performed on the KASI Science Cloud platform supported by Korea Astronomy and Space Science Institute. We acknowledge the hospitality at the Asia Pacific Center for Theoretical Physics where a part of this work was done.
\end{acknowledgements}

\bibliographystyle{aa}
\bibliography{references}


\begin{appendix}

\section{Coefficients of Complex Polynomial Lens Equation and Critical Lines for Binary Lenses}

The lens equation for the binary lens can be represented as a complex polynomial equation of degree 5:
\begin{equation}
c_5 z^5 + c_4 z^4 + c_3 z^3 + c_2 z^2 + c_1 z + c_0= 0~,
\end{equation}
for two point-like lenses, $m_1$ and $m_2$, at $z_1$ and $z_2$, respectively, in the lens plane. The coefficients $c_i$ of the polynomial equation are given as
\begin{eqnarray}
c_5 &=& -z_1^*z_2^*\!+\!(z_1^*\!+\!z_2^*)z_s^*\!-\!z_s^{*2}~, \nonumber\\
c_4 &=& m_1 z_1^*\!+\!m_2 z_2^*\!+\!(z_1^*z_2^*\!+\!z_s^{*2})(2z_1\!+\!2z_2\!+\!z_s)\!-\!z_s^*[2\mu \nonumber\\
&& +(z_1^*\!+\!z_2^*)(2z_1\!+\!2z_2\!+\!z_s)]~,  \nonumber\\
c_3 &=& -m_1(z_1z_1^*\!-\!z_1z_2^*\!+\!2z_2z_1^*)\!-m_2(z_2z_2^*\!-\!z_2z_1^*\!+\!2z_1z_2^*) \nonumber\\
&& -2\mu(z_1^*\!+\!z_2^*)z_s\!+\!2z_s^*(m_1z_2\!+\!m_2z_1\!+\!2\mu z_s)\!-\![z_1^*z_2^* \nonumber\\
&& +z_s^{*2}-(z_1^*\!+\!z_2^*)z_s^*][(z_1\!+\!z_2)^2\!+\!2z_1z_2\!+\!2(z_1\!+\!z_2)z_s]~, \nonumber\\
c_2 &=& 4\mu^2z_s\!-\!2\mu(m_1z_1\!+\!m_2z_2)\!+\!2\nu(z_1^2z_2^*\!-\!z_2^2z_1^*)\!-\!4\nu z_1z_2 \nonumber\\
&& \times (z_1^*\!-\!z_2^*)\!+\!(m_1z_1\!+\!m_2z_2\!+\!2m_1z_2\!+\!2m_2z_1)(z_1^*\!+\!z_2^*) \nonumber\\
&& \times z_s\!+\!2z_1z_2(z_1\!+\!z_2)z_1^*z_2^*\!+\!(z_1\!+\!z_2)^2z_1^*z_2^*z_s\!+\!2z_1z_2z_1^* \nonumber\\
&& \times z_2^*z_s\!+\!z_s^{*2}[2z_1z_2(z_1\!+\!z_2)\!+\!(z_1\!+\!z_2)^2z_s\!+\!2z_1z_2z_s]\nonumber\\ 
&& -z_s^*[2\nu(z_1^2\!-\!z_2^2)\!+\!2m_1(z_1\!+\!2z_2)z_s\!+\!2m_2(2z_1\!+\!z_2)z_s \nonumber\\
&& +(z_1^*\!+\!z_2^*)\{2z_1z_2(z_1+z_2)\!+\!(z_1+z_2)^2z_s\!+\!2z_1z_2z_s\}]\nonumber\\
c_1 &=& 2z_1z_2[m_1^2\!+\!m_2^2\!+\!m_1(z_2z_2^*\!+\!2z_1z_2^*\!-\!z_2z_1^*)\!+\!m_2(z_1z_1^* \nonumber\\
&& +2z_2z_1^*\!-\!z_1z_2^*)]\!+\!m_1m_2(z_1\!+\!z_2)(z_1\!+\!z_2\!-\!2z_s) \nonumber\\
&& -2[m_1^2z_2\!+\!m_2^2z_1\!-\!(z_1^*\!+\!z_2^*)\{m_1^2z_1^2\!+\!m_2^2z_2^2\!-\!2\mu \nonumber\\
&& \times(z_1\!+\!z_2)^2\}]z_s\!-\!z_1z_2z_1^*z_2^*[z_1z_2\!-\!2(z_1\!+\!z_2)]\!-\!z_1z_2z_s^{*2} \nonumber\\
&& \times[z_1z_2\!+\!2(z_1\!+\!z_2)z_s]\!-\!z_s^*[2(m_1z_1^2z_2\!+\!m_2z_1z_2^2)\!-\!2 \nonumber\\
&& \times(m_1z_2^2\!+\!m_2z_1^2\!+\!4\mu z_1z_2)z_s\!-\!z_1^2z_2^2(z_1^*\!+\!z_2^*)\!-\!2z_1z_2 \nonumber\\
&& \times(z_1\!+\!z_2)(z_1^*\!+\!z_2^*)z_s]~, \nonumber\\
c_0 &=& (m_1z_2\!+\!m_2z_1)^2z_s\!-z_1z_2(m_1^2z_2\!+\!m_2^2z_1)\!-\!m_1m_2z_1z_2 \nonumber\\
&& \times(z_1\!+\!z_2)\!-\!z_1^2z_2^2(m_1z_2^*\!+\!m_2z_1^*)\!+\!z_1z_2(m_1z_2\!+\!m_2z_1) \nonumber\\
&& \times(z_1^*\!+\!z_2^*)z_s\!+\!z_1^2z_2^2z_1^*z_2^*z_s\!+\!z_1^2z_2^2z_sz_s^{*2}\!+\!z_s^*[2\mu z_1^2z_2^2 \nonumber\\
&& -2z_1z_2(m_1z_2\!+\!m_2z_1)z_s\!-\!z_1^2z_2^2(z_1^*\!+\!z_2^*)z_s]~, 
\label{eq:le_poly_coeffs}
\end{eqnarray}
where $\mu\!=\!(m_1\!+\!m_2)/2$ and $\nu\!=\!(m_2\!-\!m_1)/2$. If we assume the masses of two lenses are the same, i.e., $z_2\!=\!-z_1$, and they are on the real axis, i.e., $z_1 = z_1^*$, the coefficients in Eq.~\eqref{eq:le_poly_coeffs} are reduced to the ones in Equation (2) of~\cite{Witt:1995}.

For the given binary lens system, the equation of the critical curves in Eq.~\eqref{eq:cri_curve} is also rewritten as a complex polynomial equation of degree 4:
\begin{equation}
d_4 z^4 + d_3 z^3 + d_2 z^2 + d_1 z + d_0 = 0~, \label{eq:cri_lines}
\end{equation}
for $\phi \in [0,2\pi)$, where
\begin{eqnarray}
d_4 &=& e^{-i\phi}~, \nonumber\\
d_3 &=& -2e^{-i\phi}(z_1\!+\!z_2)~, \nonumber\\
d_2 &=& e^{-i\phi}(z_1^{2}\!+\!4z_1z_2\!+\!z_2^{2})\!-\!(m_1\!+\!m_2)~, \nonumber\\
d_1 &=& -2e^{-i\phi}z_1z_2(z_1\!+\!z_2)\!+\!2(m_1z_2\!+\!m_2z_1)~, \nonumber\\
d_0 &=& e^{-i\phi}z_1^{2}z_2^{2}\!-\!m_1z_2^{2}\!-\!m_2z_1^{2}~. 
\label{eq:ce_poly_coeffs}
\end{eqnarray}
Then, using the lens equation, we can map the solutions of Eq.~\eqref{eq:cri_lines} to the source plane, and, eventually, get the caustics.

\end{appendix}

\end{document}